\documentclass[useAMS,usenatbib,usegraphicx]{mn2e}
  
\usepackage{aas_macros}
\usepackage{threeparttable}

\title[Supernovae in magnetized primordial halos]{Supernova explosions in magnetized, primordial dark matter halos}
  \author[D. Seifried et al.]
  {D.~Seifried,$^{1}$\thanks{dseifried@hs.uni-hamburg.de} R.~Banerjee,$^{1}$ D.~Schleicher$^{2}$ \\
  $^1$Hamburger Sternwarte, Universit\"at Hamburg, Gojenbergsweg 112, 21029 Hamburg, Germany\\
  $^2$Georg-August-Universit\"at G\"ottingen, Institut f\"ur Astrophysik, Friedrich-Hund-Platz 1, 37077 G\"ottingen, Germany
  }

\date{Released 2014}

\pagerange{\pageref{firstpage}--\pageref{lastpage}} \pubyear{2013}

\begin{document}

\label{firstpage}

\maketitle

\begin{abstract}
The first supernova explosions are potentially relevant sources for the production of the first large-scale magnetic fields. For this reason we present a set of high resolution simulations studying the effect of supernova explosions on magnetized, primordial halos. We focus on the evolution of an initially small-scale magnetic field formed during the collapse of the halo. We vary the degree of magnetization, the halo mass, and the amount of explosion energy in order to account for expected variations as well as to infer systematical dependencies of the results on initial conditions. Our simulations suggest that core collapse supernovae with an explosion energy of 10$^{51}$ erg and more violent pair instability supernovae with 10$^{53}$ erg are able to disrupt halos with masses up to a about 10$^6$ and 10$^7$ M$_{\sun}$, respectively. The peak of the magnetic field spectra shows a continuous shift towards smaller $k$-values, i.e. larger length scales, over time reaching values as low as $k$ = 4. On small scales the magnetic energy decreases at the cost of the energy on large scales resulting in a well-ordered magnetic field with a strength up to $\sim 10^{-8}$ G depending on the initial conditions. The coherence length of the magnetic field inferred from the spectra reaches values up to 250 pc in agreement with those obtained from autocorrelation functions. We find the coherence length to be as large as 50\% of the radius of the supernova bubble. Extrapolating this relation to later stages we suggest that significantly strong magnetic fields with coherence lengths as large as 1.5 kpc could be created. We discuss possible implications of our results on processes like recollapse of the halo, first galaxy formation, and the magnetization of the intergalactic medium.
\end{abstract}


\section{Introduction}

Magnetic fields play an important role in the present day universe. They pervade entire galaxies including our own Milky Way~\citep[e.g.][]{Beck99,Beck12} as well as the intergalactic medium \citep[e.g.][]{Kim89,Kronberg94}. Recent observations of TeV blazars with FERMI \citep[e.g.][]{Neronov10,Dolag11,Takahashi12} provide hints about the strength of the intergalactic magnetic field. The question, however, where these fields come from is still not conclusively answered, in particular how were the magnetic fields initially generated, and how amplified to the strength observed nowadays.

The first seed magnetic fields in the early universe could have been generated by a couple of different mechanisms. One mechanism is the generation of magnetic fields in the phase of inflation in the very early universe resulting in a field strength of $10^{-34}$ -- $10^{-10}$ G on scales of about 1 Mpc \citep{Turner88}. Another possibility is the generation via first order phase transitions, with field strengths of $B_0 = 10^{-29}$ G from the electroweak phase transition and $B_0 = 10^{-20}$ G from the QCD phase transition on scales of 10 Mpc \citep{Sigl97}. The actual coherence length, however, depends crucially on the properties of the field, like e.g. the magnetic helicity and the spectral index of the magnetic power spectrum on large scales~\citep[e.g.][]{Banerjee04}. Another process is the so-called ``Biermann battery'' \citep{Biermann50,Kulsrud08} where due to the different behaviour of electrons and ions in a plasma magnetic fields with $B_0 = 10^{-19}$ G can be generated. More recently, \citet{Schlickeiser12} showed that plasma fluctuations can generate magnetic fields of the order of $10^{-10}$ G even in a completely unmagnetized, non-relativistic medium. For comparison, in present day galaxies kpc-scale magnetic fields with strengths of a few $\mu$G are found \citep[e.g.][]{Beck99,Beck12}.

For this reason at some point a significant amplification of primordial magnetic fields on large scales has to take place. One way to accomplish this is the magnetic dynamo which amplifies the magnetic field on various scales, which is why the effects on large and small scales are usually considered separately. On large scales the so-called mean-field dynamo describes the amplification in a differentially rotating fluid in the presence of helical turbulent motions \citep[e.g.][]{Parker55,Ruzmaikin88}. On scales smaller than the driving scale of turbulence the small-scale dynamo is responsible for the amplification of the magnetic field by converting the turbulent kinetic energy into magnetic energy by stretching and twisting magnetic field lines \citep{Kazantsev68,Subramanian97,Schober12b}. In the so-called kinematic phase of the small-scale dynamo the magnetic energy experiences an exponential growth mainly at the resistivity scale. Once the magnetic field has grown to a significant strength, the non-linear phase sets in, in which the magnetic energy is transported to larger and larger scales \citep{Schleicher13}. The amplification ends once the magnetic field has reached saturation, i.e. when the magnetic energy is of the order of 10\% of the turbulent kinetic energy on the driving scale of the turbulence. 

The formation of the first stars, so called Population III stars (hereafter Pop III stars), in the universe has taken place in dark matter halos with typical masses of 10$^6$ M$_{\sun}$ \citep[e.g.][]{Bromm99,Bromm02,Abel00,Yoshida03,Gao07,OShea07}. By means of an analytical consideration, \citet{Schober12b} have shown that the small-scale dynamo can amplify an initially very weak magnetic field in such a halo to strengths of $10^{-6}$ G within short timescales, a mechanism which could also work during the formation of the first galaxies \citep{Schleicher10,Souza10,Schober13}. These analytical predictions were recently confirmed by a number of numerical simulations of Pop III star formation \citep{Sur10,Sur12,Peters12,Turk12,Latif13d,Latif13b}. The authors showed that when using a high enough spatial resolution the magnetic field in the center of the halo indeed can reach saturation.  These magnetic fields are usually small-scaled, strongly tangled, and localized in the center of the star forming halos. For this reason, the field cannot contribute to the large-scale, volume-filling magnetic field observed nowadays unless it is removed from the center and transported to larger scales.

As shown by \citet{Heger02}, Pop III stars with masses between 10 and 50 M$_{\sun}$ will end their lives with a core collapse supernova and those with masses from 140 to 260 M$_{\sun}$ with an up to 100 times more energetic pair instability supernova. The effect of supernova explosions on primordial halos was studies numerically by a number of authors \citep{Bromm03,Kitayama05,Greif07,Souza11,Ritter12,Whalen08,Whalen13b,Johnson13}. A common result of these simulations is that whether a halo is disrupted by a supernova or not crucially depends on the mass of the halo, the injection energy of the supernova, and prior to that on the ability of the progenitor star to ionize the halo and drive out gas in its inner region. Even in the case that a HII regions was created, a too small injection energy or a too high halo mass would cause the expansion of the supernova bubble to stall and the gas to start to fall back towards the center. In case the supernova is able to disrupt the halo, however, material from the center of the halo is ejected and transported to distances of the order of kpc.
So far, in these simulations the effect on magnetic fields was not studied. We note, however, that \citet{Balsara01} considered the effect of a single supernova on a magnetized medium with homogeneous density in the present day interstellar medium on scales of $\leq$ 100 pc  thus significantly smaller than the scale of interest here. The authors show that the magnetic field is swept up in the supernova shell, which can also be expected for primordial halos affected by supernovae. How exactly the magnetic field on large (kpc) scales is affected in a primordial halo will be the main focus of the work presented here.

The paper is organized as follows: In Section~\ref{sec:techniques} we describe the numerical techniques used for the simulations. In Section~\ref{sec:IC} we describe the initial conditions used for the simulated primordial halos. The results of the simulations are presented in Section~\ref{sec:results}, where we first give a general overview of the simulations before we consider the properties of the magnetic field and the effect of varying conditions. Next, the results are discussed in a broader context in Section~\ref{sec:discussion} before we summarize our work in Section~\ref{sec:conclusion}.

\section{Numerical techniques}
\label{sec:techniques}

The simulations were preformed with the versatile astrophysical code FLASH in version 4~\citep{Fryxell00,Dubey08}. We use the unsplit, staggered mesh solver, which comes along with the public FLASH version 4 \citep{Lee09}, in order to solve the magnetohydrodynamical equations under the condition of ideal magnetohydrodynamics. The solver implicitly satisfies the divergence-free constraint of the magnetic field up to the accuracy of machine round-off errors. We used a third order hydro scheme and a hybrid Riemann solver using the HLL scheme at shocks and the HLLD scheme otherwise in order to guarantee the best stability and accuracy. In all simulations the adiabatic index is set to $\gamma = 5/3$.

We do not use self-gravity in this work in order to speed up the simulations. We instead model the gravitational effect of baryonic and dark matter by assuming a spherically symmetric gravitational potential which remains constant over time. Despite the fact that the (baryonic) density distribution changes over time we consider it as a reasonable approximation for two reasons: Firstly, on the timescales considered the dark matter distribution is not expected to change significantly. Secondly, at the scales of interest ($\sim$ 10 pc -- 1 kpc) the gravitational potential is dominated by the dark matter which is why the change in the baryonic matter distribution would only have a minor effect on the overall gravitational potential. We point out that for a mean dark matter density in the halos of $200 \cdot \rho_{\rmn{crit}}$ (see equation~\ref{eq:mdm2} further below) the free-fall time is about 30 Myr thus significantly longer than the timescales considered here ($\leq$ 10 Myr), which also justifies the use of a fixed gravitational potential. We emphasize that a time-independent gravitational potential was also used by a number of other authors \citep{Kitayama04,Kitayama05,Whalen04,Whalen08,Whalen13a} despite the fact that moderate changes in the dark matter profile could be introduced by supernova explosions \citep{Souza11}.

We performed two kinds of simulations, simulations with an adiabatic behaviour ($\gamma = 5/3$), where no additional cooling is applied, and simulations using a tabulated cooling function appropriate for primordial gas with zero metallicity as a sink of thermal energy after the hydrodynamical timestep. For the latter simulations we use the cooling rates $\Lambda$ given by \citet{Sutherland93}. These include collisional excitation and ionization cooling for H and He, recombination cooling for H and He, and bremsstrahlung. We note that we are missing an accurate description for Compton cooling for a redshift of z = 20, which is why we most likely somewhat underestimate the cooling efficiency (see also Section~\ref{sec:caveats}). After each hydrodynamical timestep we update the internal energy by
\begin{equation}
 \Delta e_{int} = - \Lambda(T) \cdot n_\rmn{H}^2 \cdot \rmn{d}t
\label{eq:cooling}
\end{equation}
where $n_\rmn{H}$ is the hydrogen number density and d$t$ the hydro timestep. By using a subcycling scheme for the cooling routine we guarantee that the internal energy does not change by more than 20\% within one cooling timestep. In case more that one cooling timestep is required (since the internal energy would change by more than 20\% otherwise) the cooling rate $\Lambda$, which depends on the temperature of the gas, is updated accordingly.

\section{Initial conditions}
\label{sec:IC}

We now describe the initial conditions used in our simulations. We performed simulations with different masses of the dark matter halo, varying magnetic field strengths, and two different scaling relations between the magnetic field and the gas density. Moreover, we simulated the effect of supernova explosions with different amounts of energy. Finally, as mentioned already before we repeated some selected simulations without cooling, i.e. with a purely adiabatic behaviour in order to investigate the effect of the cooling. All simulations and their corresponding parameters are listed in Table~\ref{tab:models}. The simulations where performed in a cubic box with a side length of 1024 pc using a uniform resolution of 512$^3$ grid points, i.e. a spatial resolution of 2 pc. In order to cover a larger physical scale we have repeated one of the simulations (M3\_B1\_PISN) with a two times larger simulation box, i.e. a cube with a side length of 2048 pc. The inner region, a cube with a side length of 1024 pc, has the same spatial resolution of 2 pc as in the original run whereas in the outer parts we have a two times lower resolution of 4 pc. Hence, the grid structure is that of a nested grid with a 512$^3$ grid placed in the center of second 512$^3$ grid with a two times lower resolution.

We point out that we used highly idealized simulations described in the following to probe the influence of supernovae on the magnetic field evolution. This idealization was necessary in order to be able to cover the aforementioned parameter space with a reasonable number of simulations and to be able to use the high resolution required for a proper analysis of the magnetic field while simultaneously keeping the computational costs at a reasonable level.
\begin{table*}
 \caption{Initial conditions of the performed simulations showing the total (baryonic + dark matter) mass of the halo, the virial radius, the density of the HII region, magnetic field strength at a radius of 80 pc, the scaling relation between the magnetic field and the gas density, whether a core collapse supernova (SN) or a pair instability supernova (PISN) is used, and whether cooling is applied or not.}
 \label{tab:models}
 \begin{threeparttable}
 \begin{tabular}{@{}lccccccc}
  \hline
 Run & $M_\rmn{halo}$ & R$_\rmn{vir}$ & $n_\rmn{HII}$ & $|B|$ (80 pc) & p                    & SN/PISN & Cooling  \\
     &  [M$_{\sun}$]  & [pc]          & [cm$^{-3}$]   & [nGs] & ($B \propto \rho^{\rmn{p}}$) &         &  \\
  \hline
 M1\_B1\_SN & $1.8 \cdot 10^5$ & 78 & 0.1           &    10$^{-2}$   & 1.0                  & SN      & yes \\
 M1\_B1\_PISN & $1.8 \cdot 10^5$ & 78 & 0.1         &    10$^{-2}$   & 1.0                  & PISN    & yes \\
\hline
 M2\_B1\_SN & $7.5 \cdot 10^5$ & 129 & 0.3           &    10$^{-2}$   & 1.0                  & SN      & yes \\
 M2\_B1\_PISN & $7.5 \cdot 10^5$ & 129 & 0.3         &    10$^{-2}$   & 1.0                  & PISN    & yes \\
\hline
 M3\_B1\_noSN & $4.2 \cdot 10^6$ & 223         & 1.0           &   10$^{-2}$    & 1.0                  & none      & yes \\
 M3\_B1\_SN   & $4.2 \cdot 10^6$ & 223         & 1.0           &   10$^{-2}$    & 1.0                  & SN      & yes \\
 M3\_B1\_PISN\tnote{a} & $4.2 \cdot 10^6$ & 223         & 1.0           &    10$^{-2}$  & 1.0                  & PISN    & yes \\
 M3\_B1\_5PISN & $4.2 \cdot 10^6$ & 223        & 1.0           &    10$^{-2}$  & 1.0                  & 5 $\times$ PISN    & yes \\
\hline
 M3\_B2\_SN    & $4.2 \cdot 10^6$ & 223         & 1.0           &   1    & 1.0                  & SN      & yes \\
 M3\_B2\_PISN  & $4.2 \cdot 10^6$ & 223         & 1.0           &    1   & 1.0                  & PISN    & yes \\
 M3\_B3\_PISN  & $4.2 \cdot 10^6$ & 223         & 1.0           &    10$^{2}$   & 1.0                  & PISN    & yes \\
 M3\_B4\_PISN & $4.2 \cdot 10^6$ & 223         & 1.0           &   $1.4 \cdot 10^3$    &  0.5                  & PISN    & yes \\
\hline
 M2\_B1\_SN\_ad   & $7.5 \cdot 10^5$ & 129         & 0.3           &  10$^{-2}$   & 1.0                  & SN      & no \\
 M3\_B1\_SN\_ad   & $4.2 \cdot 10^6$ & 223         & 1.0           &  10$^{-2}$   & 1.0                  & SN      & no \\
 M3\_B1\_PISN\_ad & $4.2 \cdot 10^6$ & 223         & 1.0           &    10$^{-2}$ & 1.0                  & PISN    & no \\
 M3\_B4\_PISN\_ad & $4.2 \cdot 10^6$ & 223         & 1.0           &   $1.4 \cdot 10^3$  & 0.5                  & PISN    & no \\
  \hline
 \end{tabular}
 \begin{tablenotes}
 \item[a] This simulation was also repeated with a two times larger simulation domain.
 \end{tablenotes}
 \end{threeparttable}
\end{table*}

\subsection{Density, temperature and velocity field}
\label{sec:ICdens}

Inside a radius of 80 pc we use a flat, i.e. constant baryonic gas density. This is done in order to model the HII region created by the underlying Pop III star during its lifetime~\citep[e.g.][]{Whalen04,Kitayama04,Yoshida07,Abel07}. We use three different values for the density inside the HII region, i.e. a number density of 0.1, 0.3, 1 cm$^{-3}$. We note that a mean molecular weight $\mu$ of 1.23 typical for a primordial, metal-free atomic gas composition is used. Outside the HII region, i.e. at radii larger than 80 pc, the baryonic density profile decreases outwards with
\begin{equation}
 \rho \propto r^{-2.2}
\label{eq:dens_scaling}
\end{equation}
in accordance with cosmological simulations~\citep[e.g.][]{Abel02,Yoshida06,Greif11,Turk09,Turk12,Latif13a,Latif13b}. For the dark matter component we assume a NFW profile~\citep[][but see Section~\ref{sec:ICDM} for more details]{Navarro97}. As pointed out by several authors, Pop III stars are able to ionize dark matter halos out to kpc scales~\citep{Whalen04,Kitayama04,Yoshida07,Abel07}. For this reason we assume an initially constant temperature of 10\,000 K in the \textit{entire} halo. Although this is clearly a strong simplification, for the work presented here focussing on the interaction between the supernova remnant and the magnetic field, we consider this as a reasonable approximation.

Furthermore, we add a turbulent velocity field in the entire dark matter halo. The strength of the turbulent field is chosen such that the rms Mach number (M$_\rmn{rms}$) is of the order of 1 in agreement with recent cosmological simulations~\citep[e.g.][]{OShea08,Greif11,Latif13b}. The turbulence spectrum follows a power-law with an exponent of -5/3, i.e. Kolmogorov-like, between $k$-values from 32 up to 128 in order to model small-scale turbulent motions. We briefly note that the damping time (sound crossing time) $l/c_{\rmn{s}}$ of the fluctuations caused by the turbulence field is relatively long compared to the evolutionary time of the supernova remnants. Assuming a typical size $l$ of the fluctuations of about 1/32 (motivated by the applied turbulence spectrum) of the box size, i.e. $l$ = 32 pc, and a sound speed $c_{\rmn{s}} = 8.2$ km s$^{-1}$ for gas with 10\,000 K, we obtain a typical damping time of about 3.8 Myr (see also Section~\ref{sec:overview}).

\begin{figure}
 \includegraphics[width=\linewidth]{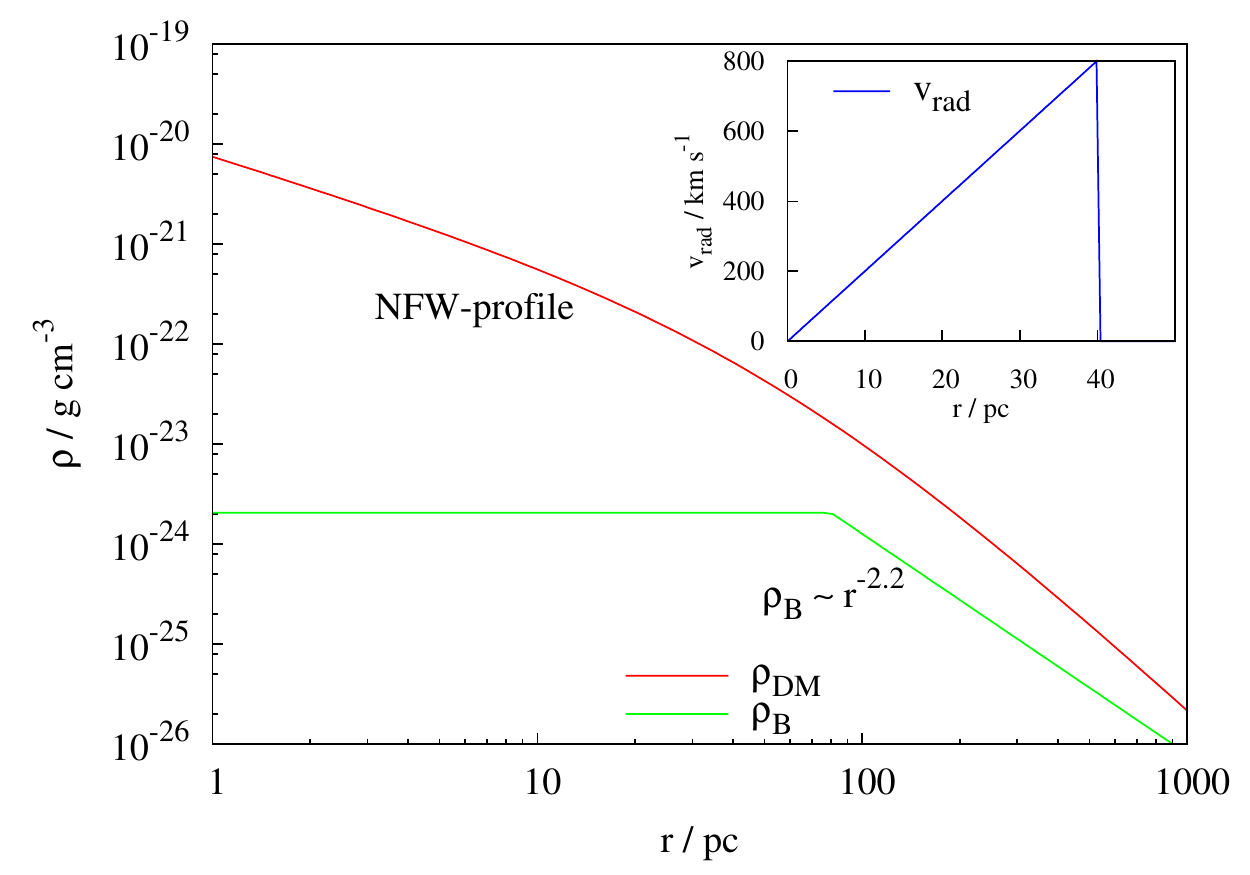}
\caption{Initial density profile for the dark matter (red line) and baryonic gas (green line) for run M3\_B1\_PISN. The inlay shows the radial velocity of the initial supernova blast wave for the same run.}
\label{fig:IC}
\end{figure}

The supernova remnant is initialized in a sphere with a radius of 40 pc thus well within the HII region. We chose a total energy E$_\rmn{SN}$ of 10$^{51}$ and 10$^{53}$ erg for the core collapse supernova and the pair instability supernova (PISN), respectively as well as an extreme case where several PISN go off simultaneously with a total energy of $5 \cdot 10^{53}$ erg. The supernova remnant is initialized within the aforementioned radius of 40 pc by depositing 70\% of the energy in form of thermal energy and the remaining 30\% as kinetic energy in form of purely radial motions\footnote{Due to the referees suggestion we repeated run M3\_B1\_PISN by injecting the entire energy in form of kinetic energy, setting the gas energy in the remnant to 10\,000 K. In particular at later times, the results do not differ significantly, thus the main conclusions of this work would not be affected.} For the sake of simplicity, the initial density inside the remnant is kept constant having the same value as in the surrounding HII region. The velocity of the blast wave increases linear with the radius in order to mimic the velocity profile of a Sedov-Taylor solution. The slope of the velocity profile is adapted such that the kinetic energy sums up to the required kinetic explosion energy. For demonstrative purposes, in Fig.~\ref{fig:IC} we show the initial density profile for the dark matter and baryonic gas as well as the radial velocity of the initial supernova blast wave for run M3\_B1\_PISN.

\subsection{Magnetic field}
\label{sec:ICbfield}

Little is known about the magnetic field strength in the first star-forming dark matter halos so far. There exist only a few cosmological simulations including magnetic fields following the formation of dark matter halos down to pc scales~\citep{Turk12,Latif13d,Latif13b}. For this reason we have little constraints on possible magnetic field strengths during the formation of the first stars, which is why we explore a large range of possible values. However, the aforementioned simulations as well as related work on the dynamo effect suggest that before \citep{Wagstaff13} and during the formation and lifetime of Pop III stars \citep{Sur10,Sur12,Schober12b} significant magnetic fields might have been present. We take the results of~\citet{Turk12} and \citet{Latif13d,Latif13b} to motivate our choice of the initial magnetic field in our simulations. Their simulation show that the magnetic field in star forming halos usually scales approximately linear with the density. Given the scaling of the density in equation~\ref{eq:dens_scaling}, this results in a scaling of the magnetic field as
\begin{equation}
 B \propto \rho \rightarrow B \propto r^{-2.2} \, .
\label{eq:scaling}
\end{equation}
For the gross of our simulations we use this scaling relation (see Table~\ref{tab:models}). In Table~\ref{tab:models} we give the mean strength of the magnetic field at the boundary of the HII region, i.e. at $r = 80$ pc. Due to the large uncertainties we explore magnetic field strengths ranging over 4 orders of magnitude. We emphasize, however, that in every case the magnetic field is not dominant, i.e. the ratio of the magnetic to thermal pressure is significantly below unity everywhere.

Since we are interested in the effect of supernova explosions on a small-scale, strongly tangled magnetic field, the values for the strength of B given in the fifth column of Table~\ref{tab:models} are radially averaged. In order to create such a small-scale, strongly disordered magnetic field, we first generate a turbulent, vector field whose spectrum follows a power-law with an exponent of -5/3 between $k$ = 32 and $k$ = 128, and has a steep drop-off outside this range. Hence, we guarantee the magnetic field to have a small coherence lengths. In order to achieve the scaling relation given in equation~\ref{eq:scaling}, in a second step we scale this vector field with the radius according to equation~\ref{eq:scaling} and finally take the curl of it. We emphasize that this method guarantees that the magnetic field is indeed divergence free despite its turbulent nature.

Finally, we note that in dynamo theory the value of the saturated magnetic field only depends on the strength of the turbulence with the magnetic energy accounting for about 10\% of the kinetic energy~\citep{Kazantsev68,Schober12a}. Given the constant velocity dispersion in our simulations (M$_\rmn{rms} \simeq$ 1), in case of saturation this would imply a scaling of the magnetic field strength with the square root of the density. Taking into account the scaling relation of equation~\ref{eq:dens_scaling} we thus would have
\begin{equation}
 B \propto \rho^{0.5} \rightarrow B \propto r^{-1.1} \, .
\label{eq:scaling2}
\end{equation}
Hence, in order to model a saturated magnetic field we performed one simulations with the above scaling relation with the strength adapted such that the ratio of the magnetic to turbulent kinetic is 0.1. Here we stress that so far a saturated magnetic field over the \textit{entire} extension of a dark matter halo was not yet observed in cosmological simulations and that it is highly speculative whether the magnetic field in the outer parts of the halo will ever reach saturation during the lifetime of the central Pop III star. Nevertheless, since this field configuration present an upper limit of the strength a potential magnetic field could have, we have chosen this particular setup considering it as an interesting numerical experiment.

\subsection{Dark matter profile}
\label{sec:ICDM}

We assume that throughout the simulation the dark matter distribution remains unchanged and follows a NFW profile~\citep{Navarro97}
\begin{equation}
 \rho_\rmn{DM}(r) \propto \frac{1}{\frac{r}{R_s}\left(1+\frac{r}{R_s}\right)^2} \, ,
\label{eq:NFW}
\end{equation}
where $R_s$ is the scaling radius which we have chosen to set to 4 times the virial radius $R_\rmn{vir}$. We scale the dark matter density profile such that inside $R_\rmn{vir}$ the mass of dark matter $M_\rmn{DM}$ is
\begin{equation}
 M_\rmn{DM} = \frac{\Omega_\rmn{DM}}{\Omega_\rmn{B}} \cdot M_\rmn{B} \, ,
\label{eq:mdm1}
\end{equation}
with $\Omega_\rmn{DM} = 0.26$ and $\Omega_\rmn{B} = 0.04$ being the dark matter and baryonic density in $\Lambda$CDM cosmology notation. Furthermore, it is known from the top-hat collapse solution~\citep{Eke96,Navarro97} that inside the virial radius the dark matter mass is
\begin{equation}
 M_\rmn{DM} = \frac{4 \pi}{3} \cdot 200 \cdot \rho_\rmn{crit} \cdot R_\rmn{vir}^3 \, ,
\label{eq:mdm2}
\end{equation}
where $\rho_\rmn{crit}$ is the critical density of the universe at z = 20. Combining equation~\ref{eq:mdm1} and~\ref{eq:mdm2} we can solve for $R_\rmn{vir}$ which in turn allows us to determine the scaling factor of the NFW profile (equation~\ref{eq:NFW}). The total (baryonic + dark matter) masses of the three different halos considered here range from $1.8 \cdot 10^5$ to $4.2 \cdot 10^6$ M$_{\sun}$ thus spanning the typical range of halo masses during Pop III star formation~\citep[e.g.][]{Bromm99,Bromm02,Abel00,Yoshida03,Gao07,OShea07}. We again point out that throughout the simulation the gravitational potential determined from the dark matter and (initial) baryonic component is kept fixed as already done in a number of other studies \citep{Kitayama04,Kitayama05,Whalen04,Whalen08,Whalen13a}.

\section{Results}
\label{sec:results}

In this section, we present the results of the simulations focussing mainly on the properties of the magnetic field. First, however, we study the general evolution of the supernova remnant over time. Subsequently, we analyse the time evolution of the magnetic field for two fiducial simulations M3\_B1\_SN and M3\_B1\_PISN, which in our case only differ by the amount of supernova energy injected. Finally, we study the effect of varying initial conditions as well as switching off the gas cooling.

\subsection{Overview}
\label{sec:overview}

\begin{figure*}
 \includegraphics[width=0.32\linewidth]{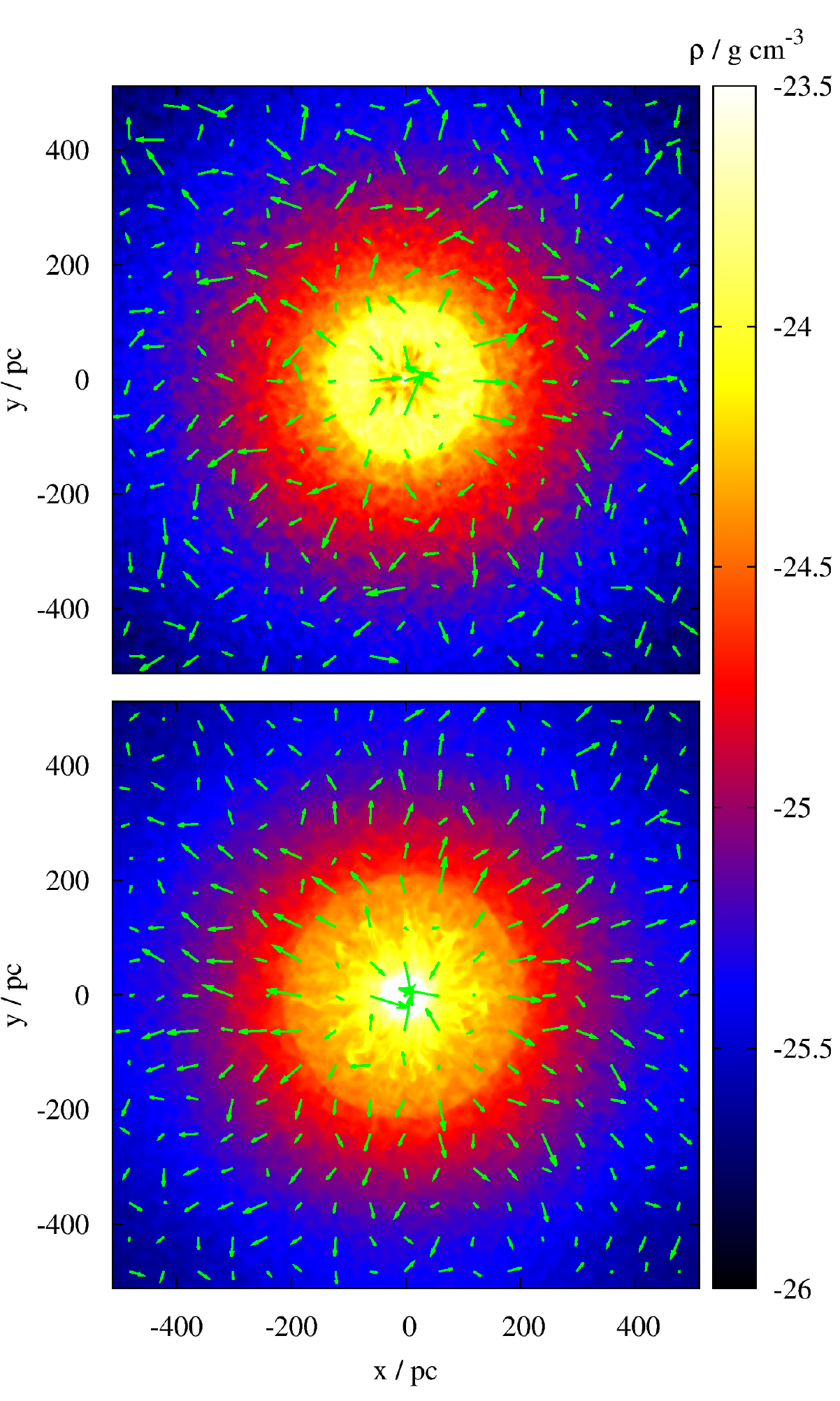}
 \includegraphics[width=0.32\linewidth]{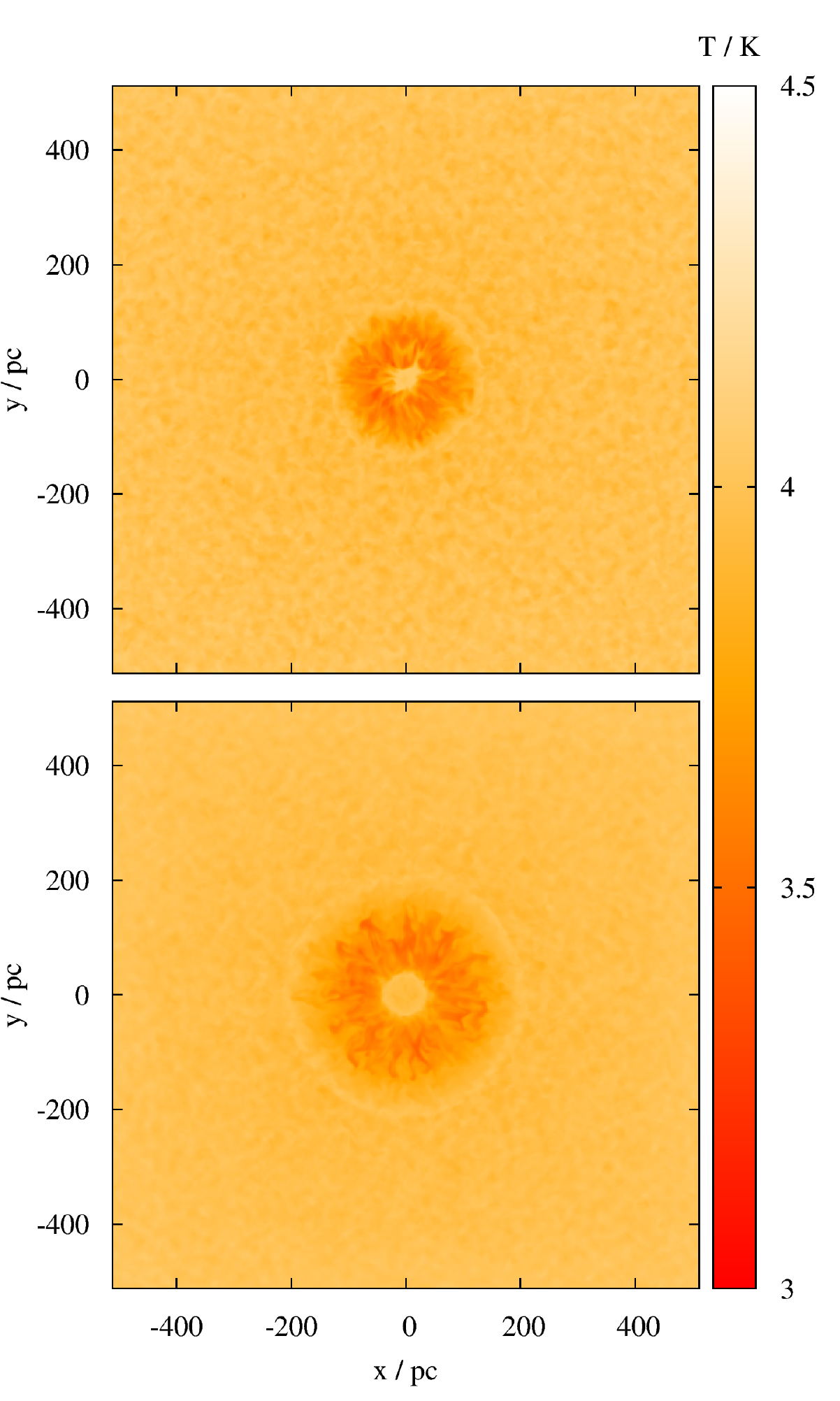}
 \includegraphics[width=0.32\linewidth]{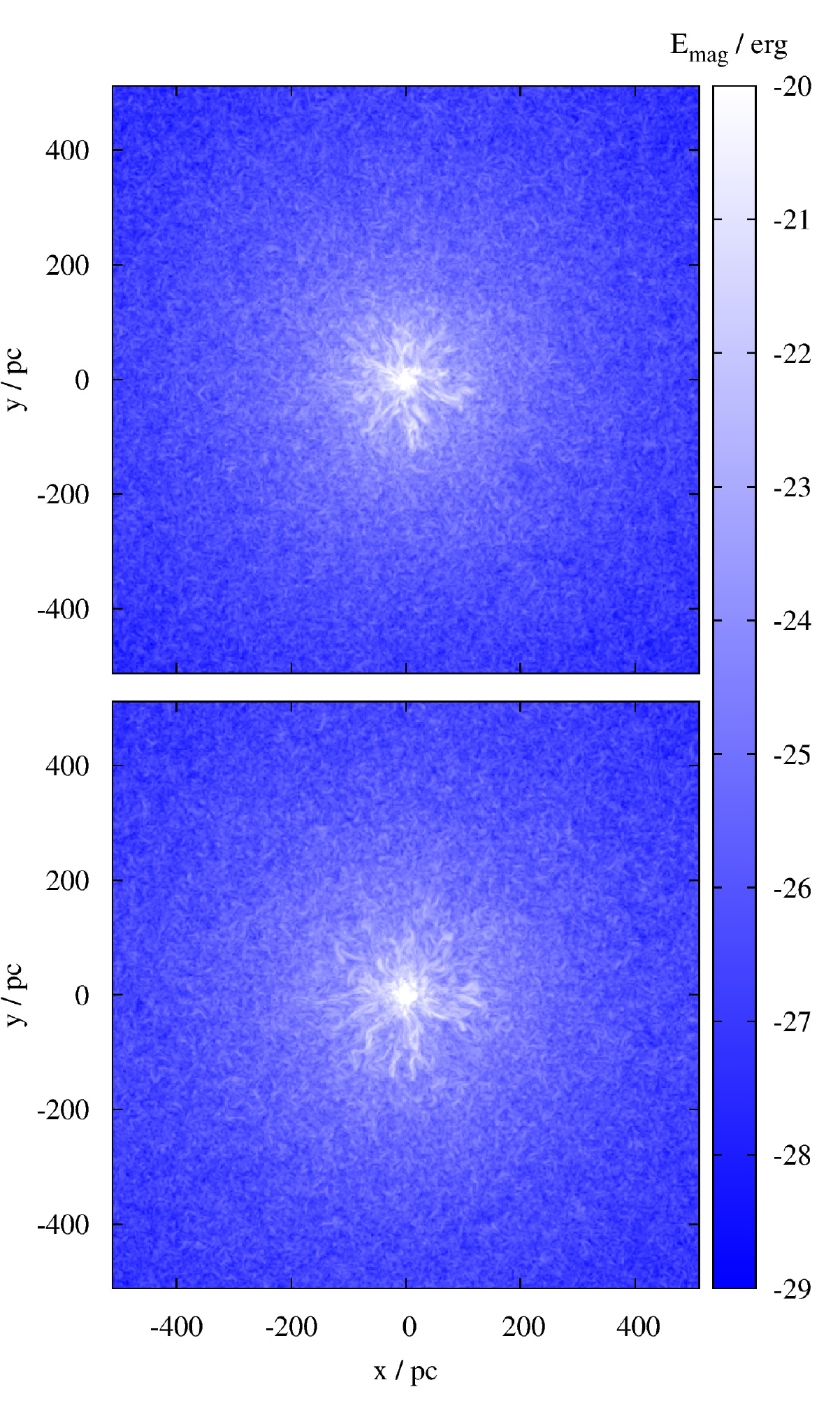}
\caption{Slices through the center of the simulation domain for run M3\_B1\_SN after 5 Myr (top) and 10 Myr (bottom). Shown are the density and velocity field (left panel), the temperature (middle panel), and the magnetic energy (right panel).}
\label{fig:slicesSN51}
\end{figure*}

\begin{figure*}
 \includegraphics[width=0.32\linewidth]{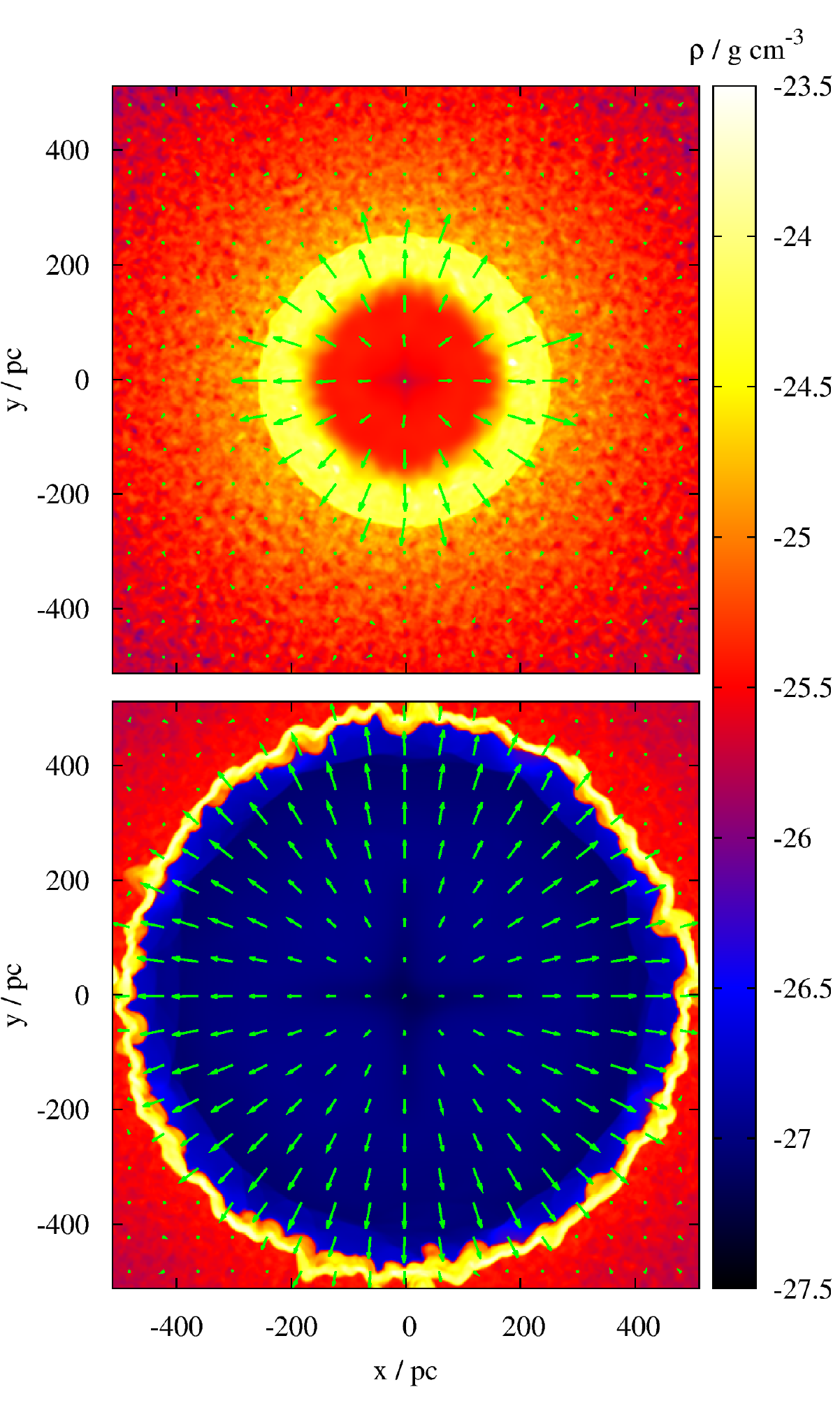}
 \includegraphics[width=0.32\linewidth]{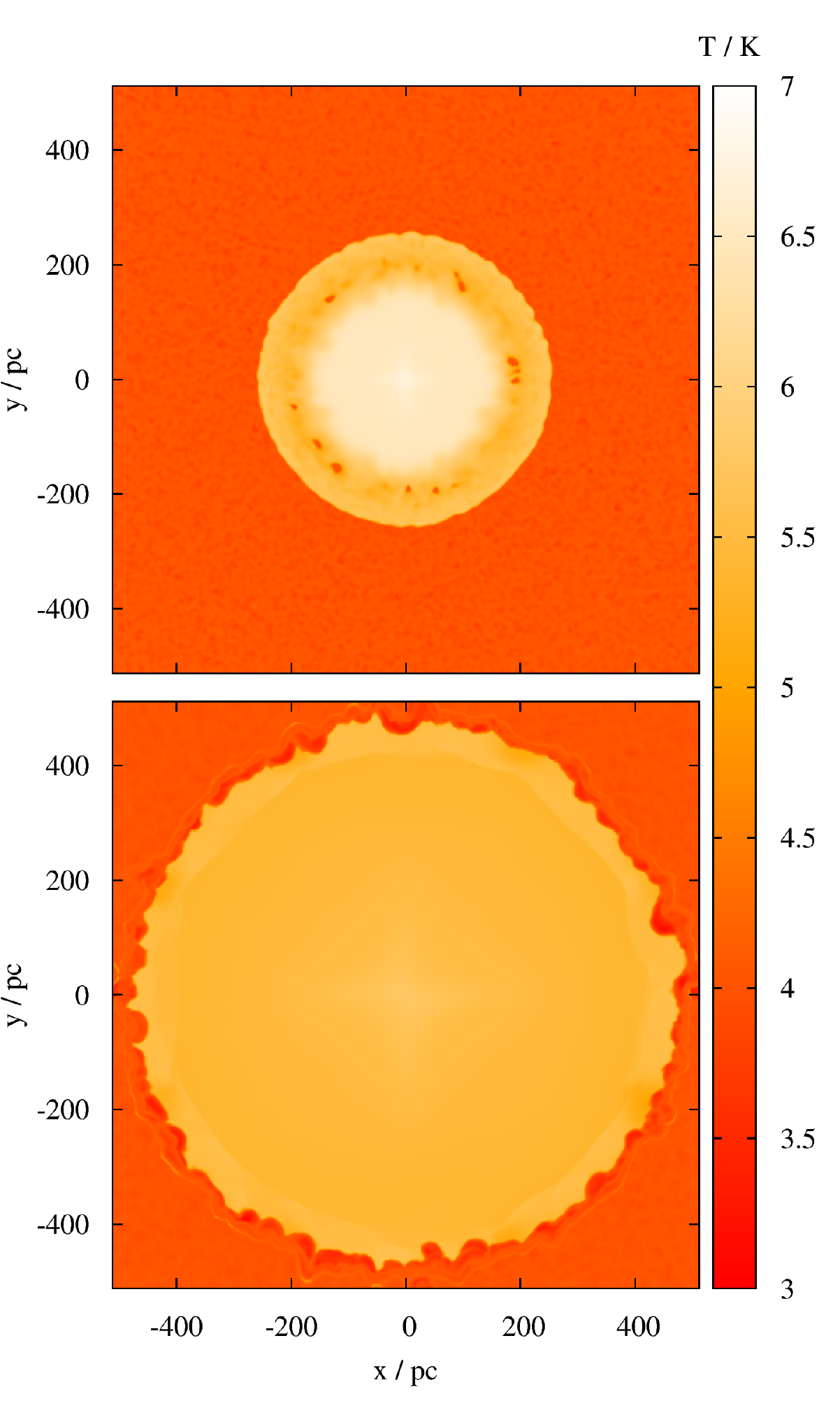}
 \includegraphics[width=0.32\linewidth]{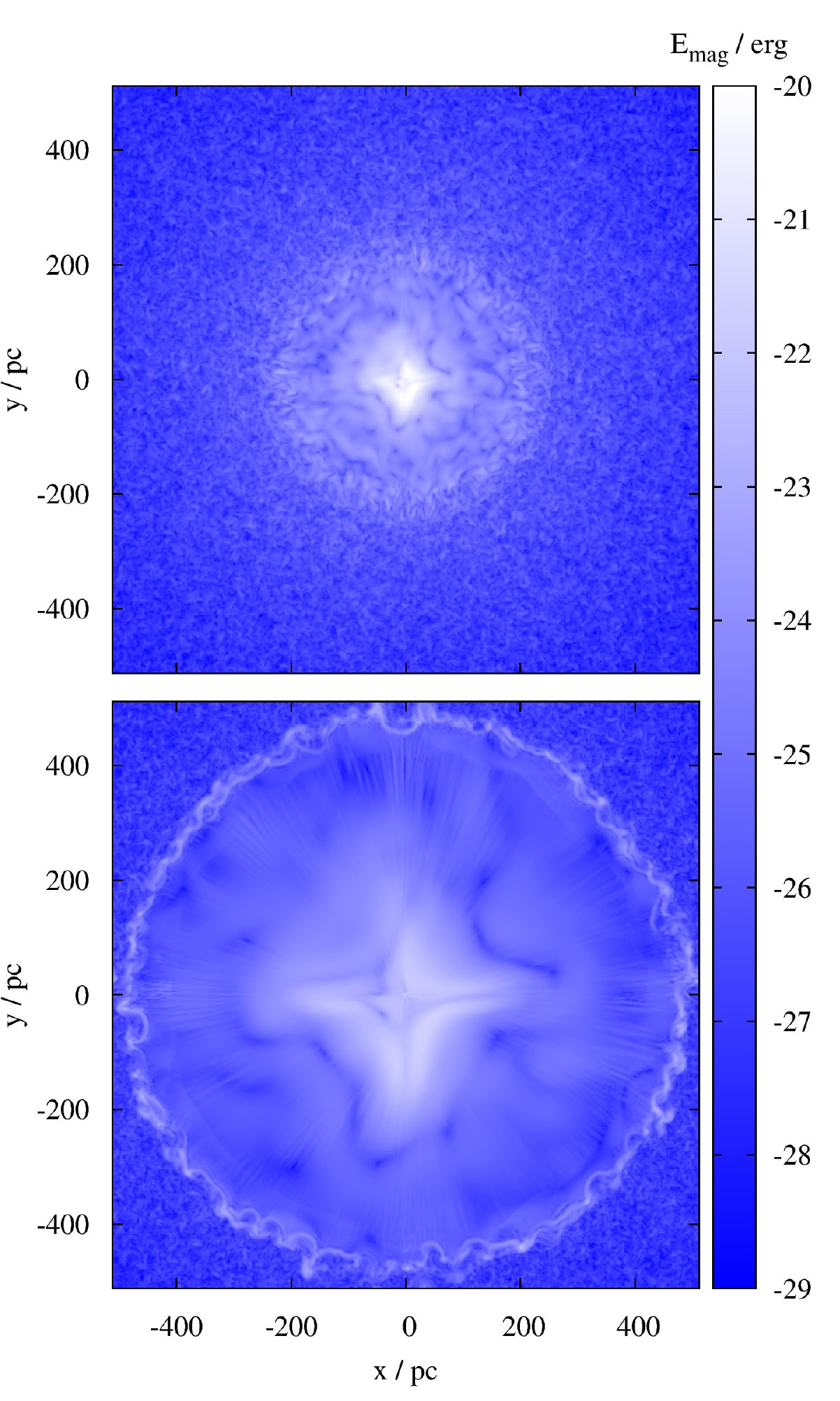}
\caption{Slices through the center of the simulation domain for run M3\_B1\_PISN after 0.96 Myr, when the diameter of the remnant is equal to half the box size (top), and after 4.43 Myr, when the diameter is equal to the box size (bottom). Shown are the density and velocity field (left panel), the temperature (middle panel), and the magnetic energy (right panel).}
\label{fig:slicesSN53}
\end{figure*}

In Fig.~\ref{fig:slicesSN51} and~\ref{fig:slicesSN53} we plot slices through the center of the simulation domain for our fiducial runs M3\_B1\_SN and M3\_B1\_PISN, respectively. The snapshots are taken at two different points in time showing the density, velocity, temperature, and magnetic field energy. We note that the snapshots are not taken at the same physical times. For run M3\_B1\_PISN (Fig.~\ref{fig:slicesSN53}) the snapshots are taken once the diameter of the PISN is equal to half the box size (0.96 Myr) and 1 times the box size (4.43 Myr), respectively. For run M3\_B1\_SN (Fig.~\ref{fig:slicesSN51}), where the supernova shock front does not reach the boundaries of the simulation domain, the snapshots are taken at 5 and 10 Myr.

As can be seen in Fig.~\ref{fig:slicesSN51}, the core collapse supernova (E$_\rmn{SN} = 10^{51}$ erg) has not managed to escape the dark matter halo after 10 Myr. In contrast, inside the shock front, which is still slowly expanding, the gas starts to recollapse towards the center resulting in an increased density towards the center. We note that the slight expansion motions in front of the supernova shock are due to the initially high gas temperature of 10000 K in the halo. This overall expansion, however, does not significantly affect the magnetic field structure as we have shown by means of a simulation without a supernova explosion (see next section). The temperature inside the supernova remnant (middle panel of Fig.~\ref{fig:slicesSN51}) drops significantly over the 10 Myr due to the applied cooling, from initially about $40 \cdot 10^6$ K to several 10$^3$ K at the end of the simulation.

In contrast to the core collapse supernova the PISN managed to entirely disrupt the baryonic component of the halo with the shock front reaching the boundary of the simulation domain (Fig.~\ref{fig:slicesSN53}). No signs of a recollapse inside the shock front are observable. The temperature of the gas in the remnant remains significantly higher than in the outer parts. This is due to the fact that 100 times more thermal energy was available and that the density in the remnant is much lower significantly decreasing the cooling ability of the gas (see equation~\ref{eq:cooling}). The radial velocity and density are lowest in the center and increase outwards whereas the temperature is highest in the center. The magnetic field reveals a significant change in it structure. As can be seen already by eye, the magnetic field appears to be more ordered inside the shock front. Furthermore, the shell exhibits a strong magnetic field, which was swept up during the expansion, an effect already observed by \citet{Balsara01} in an uniform medium.

The instabilities seen at the shock front in the bottom left panel of Fig.~\ref{fig:slicesSN53} are most likely so-called Vishniac instabilities occurring for spherical blast waves \citep{Vishniac83}. In a reference run without any turbulent motions (not shown here), we can see that these instabilities do not show up. Hence, the shock front instabilities in run M3\_B1\_PISN are triggered by the moderate density fluctuations caused by the transonic turbulence field in the ambient medium. As discussed in Section~\ref{sec:ICdens}, these fluctuations persist over the entire evolutionary time of the supernova remnant and thus will also affect its evolution at late stages. We point out that the tangled magnetic field, which was included in the reference run without turbulent motions, is not able to cause any significant density perturbations or shock front instabilities. This can be easily understood by comparing the magnetic pressure to the thermal pressure. Since the magnetic pressure is significantly smaller than the thermal pressure for all of our runs, fluctuations caused by the magnetic field will be largely suppressed or damped out very quickly. Some moderate instabilities are also present at and behind the shock front at early stages already (top left panel of Fig.~\ref{fig:slicesSN53}). However, due to our limited resolution of 2 pc, the shock fronts are not well resolved in our runs which is why we cannot follow in detail the growth of e.g. Rayleigh-Taylor instabilities \citep[e.g.][]{Chevalier92}.

The above described results are representative for all runs with a total halo mass of $4.2 \cdot 10^6$ M$_{\sun}$ (runs beginning with ``M3''). In these runs the remnant of the core collapse supernova (E$_\rmn{SN} 10^{51}$ erg) starts to recollapse in the center after a few Myr with the outer shock reaching a distance of $\sim$ 200 pc after 10 Myr. In contrast, the PISN leads to a complete disruption of the halo with no signs of recollapse. For the runs with the lowest halo mass (runs beginning with ``M1''), even the remnant of the core collapse supernova keeps expanding (run M1\_B1\_SN) without any sign of fallback although here after 10 Myr the shock has reached a distance of $\sim$ 400 pc only, thus well off from the boundary of the simulation domain. For the runs with an intermediate halo mass of $7.5 \cdot 10^5$ M$_{\sun}$, the core collapse supernova starts to recollapse in the center whereas the outer parts keep expanding after 10 Myr (see also Section~\ref{sec:disruption}).

\begin{figure}
 \includegraphics[width=\linewidth]{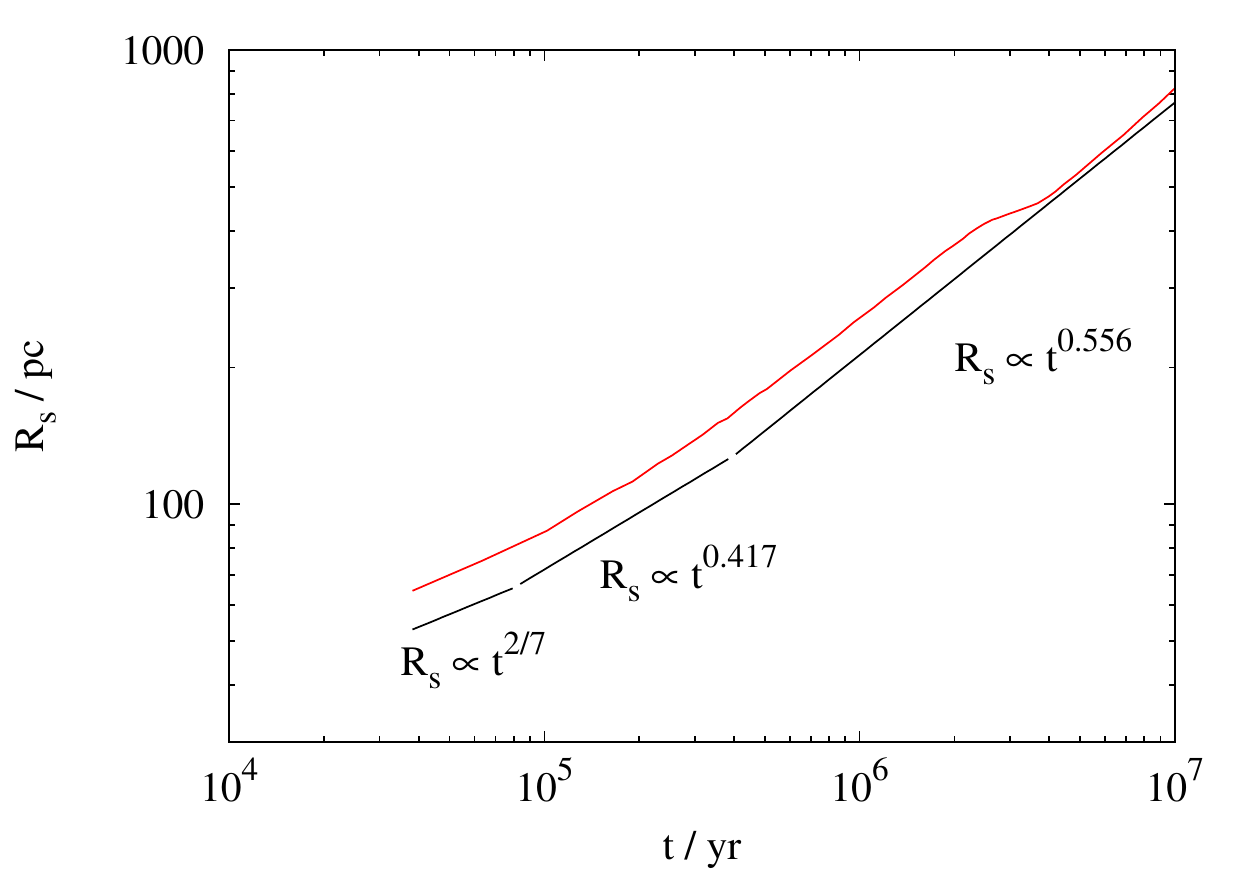}
\caption{Time evolution of the radius of the supernova shell for run M3\_B1\_PISN. Also shown are the analytical scaling relations for the PDS and MCS phase (see text).}
\label{fig:Rshell}
\end{figure}

Finally we show the position of the shock $R_\rmn{s}$ as a function of time for run M3\_B1\_PISN in Fig.~\ref{fig:Rshell}. We have used the data from the simulation with the two times larger simulation domain. We followed this simulation over 10 Myr since the shock front has not reached the boundaries of the domain by this time. As can be seen in Fig.~\ref{fig:Rshell}, by the end of the simulation the shock has travelled a distance of about 830 pc. In addition to the shock position we show the analytically derived power law relation for the position of the shock with time, $R_\rmn{s} \propto t^{\eta}$. We have chosen $\eta = 2/7$, 0.417, and 0.556, respectively. These power law exponents correspond to the theoretical prediction for the pressure-driven snowplow (PDS) phase for a supernova blast wave in a homogeneous medium ($\eta = 2/7$), and for the PDS phase and the momentum conserving (MCS) phase in a medium with the baryonic density scaling as $\rho \propto r^{-2.2}$. As shown by~\citet{Ostriker88}, in such a stratified medium the exponent for the PDS phase is $\eta = \frac{2}{7-2.2} \simeq 0.417$, and for the MCS phase $\eta = \frac{1}{4-2.2} \simeq 0.556$\footnote{See their equation 6.14 for the PDS phase and 6.13 for the MCS phase with $\gamma$ set to 5/3 and $k_{\rho}$, the density-scaling exponent, set to 2.2.}. In general, the observed time evolution of the position of the shock agrees reasonably well with the theoretical predictions. The transition from the PDS to the MCS phase seems to occur around 400 kyr which is in reasonable agreement with the result of~\citet{Greif07} finding the transition to occur around 10$^6$ yr. The difference to the work of~\citet{Greif07} might be explained by the different thermodynamical description used here. We note that there is a slight deviation from the scaling relation at late times in the MCS phase ($t \sim 3$ Myr), which we attribute to the occurrence of instabilities in the shock front. However, towards the very end it seems that the scaling relation expected from theory for the MCS phase reestablishes again.

We note that the scaling exponent of the Sedov-Taylor (ST) phase in a stratified medium with $\rho \propto r^{-2.2}$ would be $\eta \simeq 0.71$. Such a scaling relation clearly does not show up here. For shell radii smaller than 80 pc -- where the density is still homogeneous -- the scaling relation for the ST phase would exhibit the well known exponent $\eta = 2/5$ which is basically indistinguishable from the exponent $\eta = 0.417$ of the PDS phase for the stratified medium. However, it seems that in this range the curve is fitted better with the scaling relation for the PDS phase ($\eta = 2/7$). We briefly note that our scaling relation somewhat differs from that of \citet{Whalen08} (see their Fig. 10). However, since our findings fit very well with the theoretical predictions, we will not further pursue this difference here.

\subsection{Spectra}
\label{sec:spectra}

\begin{figure*}
 \includegraphics[width=0.48\linewidth]{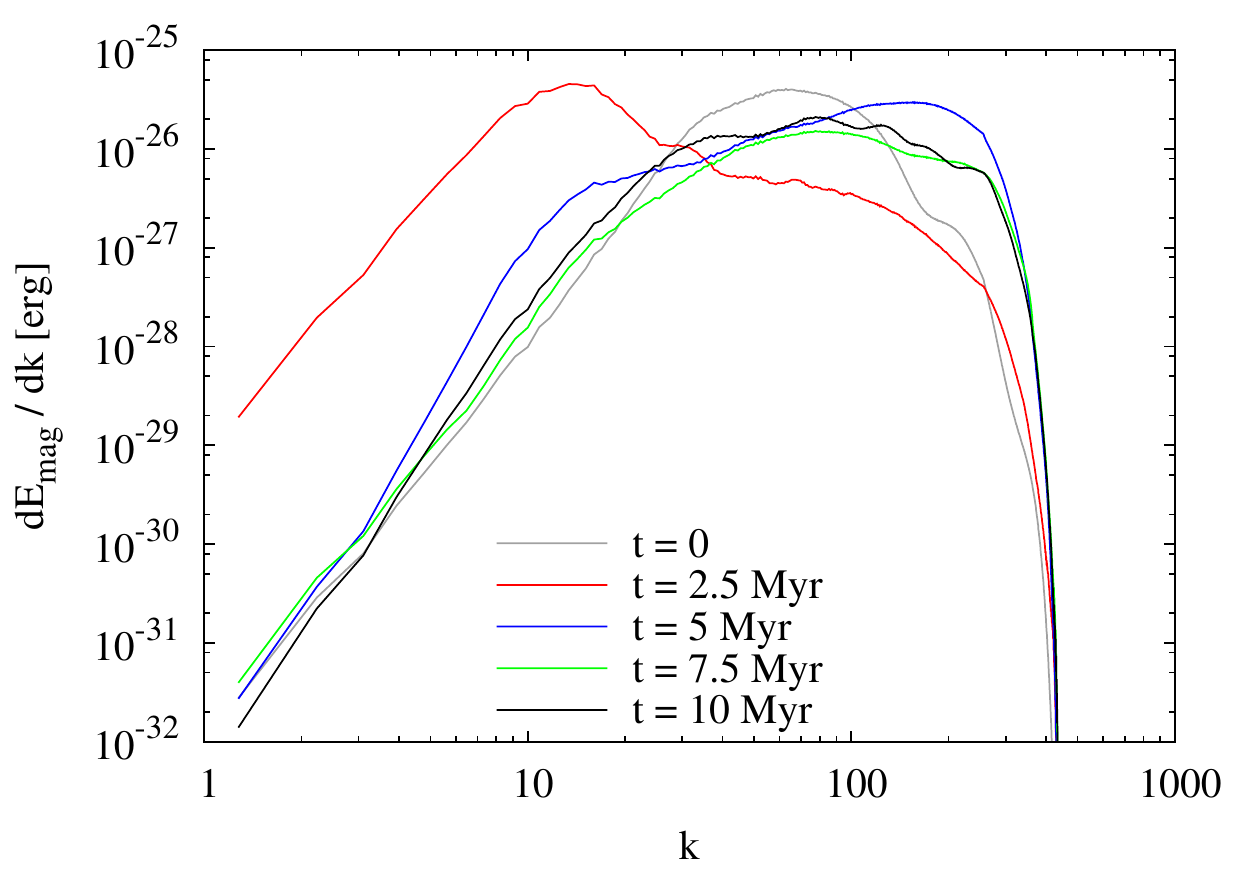}
 \includegraphics[width=0.48\linewidth]{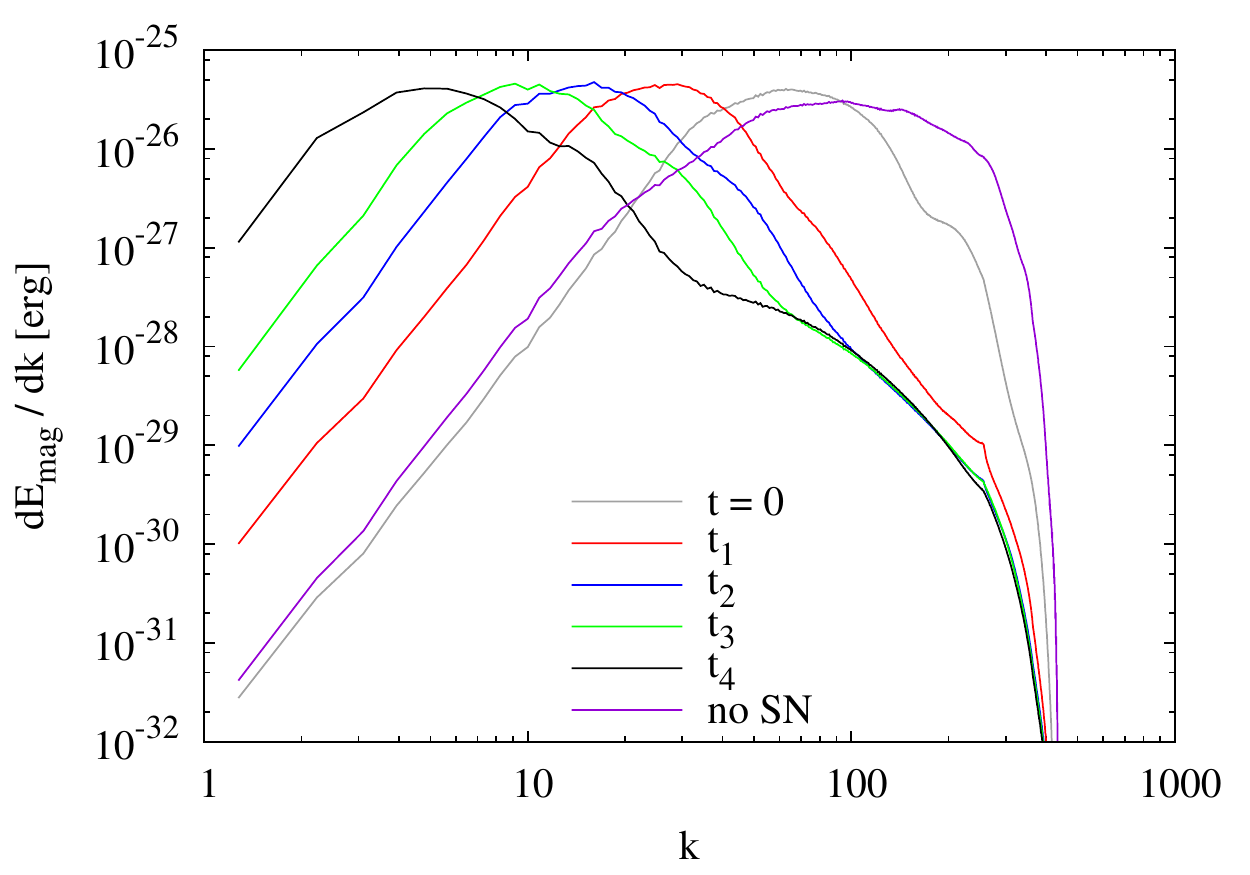}
\caption{Magnetic field spectra of the runs M3\_B1\_SN (left) and M3\_B1\_PISN (right). The times $t_1$, $t_2$, $t_3$, and $t_4$ in the right panel correspond to times where $2 \cdot R_\rmn{s}$ = 0.25, 0.5, 0.75, and 1 $\cdot$ L$_\rmn{Box}$.}
\label{fig:spectrum}
\end{figure*}

Next, we analyse the time evolution of the magnetic field spectrum in the fiducial simulations M3\_B1\_SN and M3\_B1\_PISN. As mentioned already in Section~\ref{sec:ICbfield}, the initial magnetic field is strongly tangled on small scales. In Fig.~\ref{fig:spectrum} we show the magnetic field spectra for both simulations for five different points in time. For run M3\_B1\_SN (left panel) we show the spectra at $t$ = 0, 2.5, 5, 7.5 and 10 Myr, respectively, and for run M3\_B1\_PISN (right panel) at $t$ = 0 and once the remnant has reached a diameter of 0.25 ($t_1$), 0.5 ($t_2$), 0.75 ($t_3$) and 1 ($t_4$) times the size of the computational domain\footnote{This corresponds to times of 0.25, 0.96, 2.05, and 4.43 Myr.}.
Initially the spectra peak around a $k$-value of about 60 -- 70\footnote{We point out that the initial spectrum does not follow a power-law between $k$ = 32 and 128 since the initial vector field was weighted with the radial distance as given in equation~\ref{eq:scaling}.}. We note that throughout the paper we take the wave vector $k$ as a dimensionless number measured with respect to the size of the computational domain of 1024 pc.

As can be seen, for run M3\_B1\_SN the peak of the spectrum first moves to smaller values reaching a minimum $k$-value of about 15 at $t$ = 2.5 Myr. However, after this initial phase the spectrum is again shifted to higher $k$-values peaking around $k$ = 80 at $t$ = 10 Myr. This can be explained by the occurrence of the recollapse of the remnant resulting in a compression of the gas in the center and an according increase of the magnetic field strength at smaller scales, which can also be seen in Fig.~\ref{fig:slicesSN51}. For comparative purposes, in the right panel of Fig.~\ref{fig:spectrum} we included the spectrum of the magnetic field in a reference simulation of a halo without any supernova (violet line). The spectrum is taken at $t$ = 10 Myr and is similar to the final spectrum of run M3\_B1\_SN despite the initial phase of expansion in the latter case. It is, however, significantly different to the spectra of run M3\_B1\_PISN we discuss next.

For run M3\_B1\_PISN the remnant keeps expanding which is also reflected in the evolution of the magnetic field spectrum. The peak of the spectrum is shifted to lower and lower $k$-values reaching k$_\rmn{peak} \simeq 5$ once the shock reaches the boundary of the simulation domain. Hence, at the end of the simulation the typical coherence length for the magnetic field is about 1 order of magnitude higher than for the initial configuration ($k_\rmn{peak} \simeq 60 - 70$). Our simulations therefore show that the PISN explosion has a strongly ordering effect on the magnetic field increasing the typical coherence length by about a factor of 10 with the peak of the spectrum being close to the theoretical limit of k$_\rmn{peak}$ = 2. Interestingly, the shift of the spectrum occurs in an almost self-similar manner, in particular the height of the peak does not change significantly over time. Hence, it seems that the magnetic energy stored at the largest scales (the peak scale) does not suffer any significant losses. The magnetic field on large-scales rather builds up at the cost of the small-scale fluctuations as indicated by the continuous decrease of the magnetic energy at high $k$-values over time.

\begin{figure}
 \includegraphics[width=\linewidth]{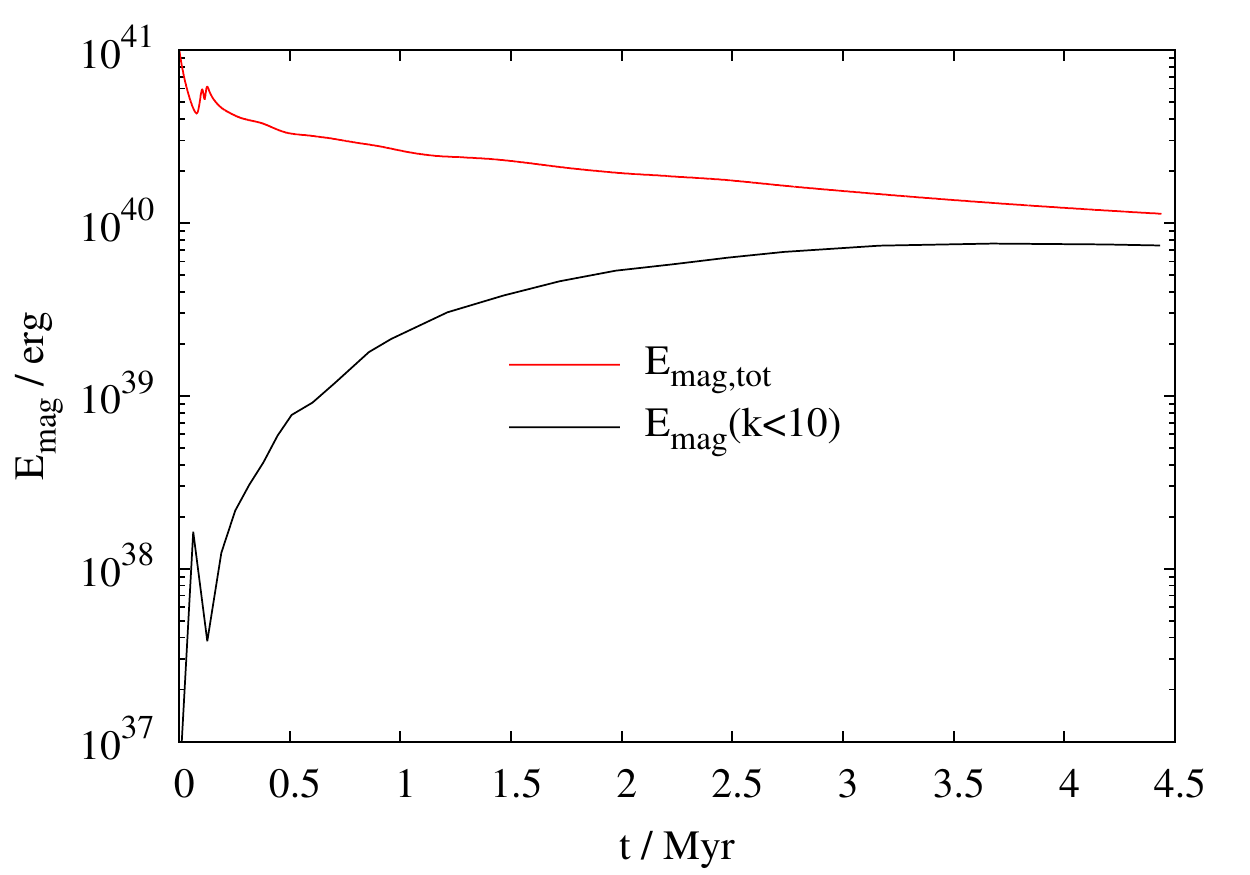}
\caption{Time evolution of the total magnetic field energy and the energy stored on large scales (see text) for run M3\_B1\_PISN.}
\label{fig:bfield}
\end{figure}

In order to further demonstrate this, in Fig.~\ref{fig:bfield} we plot the time evolution of the total magnetic energy E$_\rmn{mag,tot}$ and the energy stored on large scales E$_\rmn{mag}$(k$<$10). For the calculation of E$_\rmn{mag}$(k$<$10) we integrate the magnetic field spectra of run M3\_B1\_PISN at different times up to a $k$-value of 10. First it can be seen that the total magnetic energy decreases (almost constantly) over time, which we attribute to the loss of magnetic energy stored on the smallest scales as indicated in the right panel of Fig.~\ref{fig:spectrum}. We suppose that this overall loss is due to numerical dissipation of the small-scale fluctuations getting compressed in the thin shock layer. We note that the double peak seen in the very beginning is due to the initial refraction wave travelling inside the shock front. The energy on large scales, however, shows a continuous increase over time accounting for about 65\% of the magnetic energy at the end of the simulation. This clearly demonstrates the ordering effect of the supernova and the redistribution of the initially strongly centralized magnetic energy to larger scales (compare right panel of Fig.~\ref{fig:slicesSN53}). In summary, the PISN therefore seems to provide an efficient way to shift magnetic energy from small to large scales building up a significant large-scale magnetic field.

\subsection{Structure function}
\label{sec:SF}

The spectra shown in the section before are always convolved with the adopted scaling relation of the magnetic field (equation~\ref{eq:scaling}). Moreover, the magnetic field structure outside the shock front is also taken into account for the calculation of the spectra. A simple way to avoid the convolution of the magnetic field analysis with the initial scaling relation and regions outside the shock front is the calculation of the autocorrelation function of the magnetic field within the region enclosed by the shock front. The autocorrelation function of the magnetic field is defined as
\begin{equation}
 f(|\vec{r}|) = \left< \frac{\vec{B}(\vec{x})}{|\vec{B}(\vec{x})|} \cdot \frac{\vec{B}(\vec{x}+\vec{r})}{|\vec{B}(\vec{x}+\vec{r})|} \right>_{\vec{x}} \, ,
\label{eq:auto}
\end{equation}
where $\left< \right>_{\vec{x}}$ denotes the spatial average over a certain region. The function $f(r)$ describes the mean of the cosine of the angle $\theta$ between the magnetic field at two points separated by a distance of $r$, i.e.
\begin{equation}
 f(r) = \left< (cos \, \theta) \, (r) \right >_{\vec{x}} \; , \; \theta = \angle (\vec{B}(\vec{x}),\vec{B}(\vec{x}+\vec{r})) \, .
\end{equation}
We note that for a random vector field the structure function would be identical to zero everywhere.

\begin{figure}
 \includegraphics[width=\linewidth]{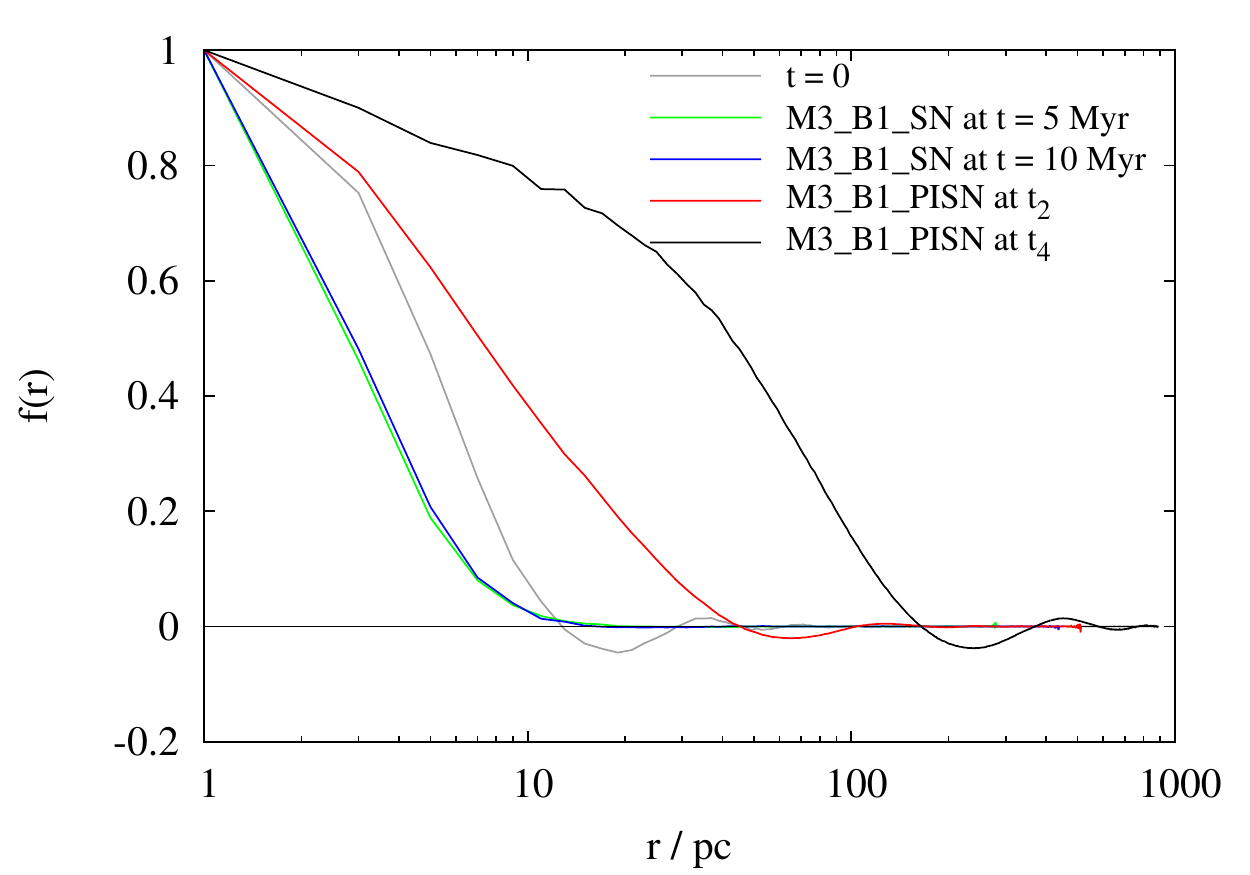}
\caption{Autocorrelation function as given by equation~\ref{eq:auto} for the runs M3\_B1\_SN and M3\_B1\_PISN at the beginning (grey line) and two later times (see text for details).}
\label{fig:auto}
\end{figure}

As mentioned before, we calculate the autocorrelation function only for pairs of points \textit{within} the supernova remnant. For the results shown in the following we have evaluated $f(r)$ for about $4\cdot 10^9$ randomly chosen pairs in order to reduce the statistical fluctuations. The autocorrelation functions for the runs M3\_B1\_SN and M3\_B1\_PISN are shown in Fig.~\ref{fig:auto} for three different points in time. For run M3\_B1\_SN the autocorrelation function is shown at $t$ = 0, 5, and 10 Myr (left panel) and for run M3\_B1\_PISN at $t$ = 0, and once the remnant has a size of 0.5 ($t_2$) and 1 ($t_4$) times the box size\footnote{Note that at $t$ = 0 the autocorrelation function is identical for both runs.}. As can be seen, initially the autocorrelation function is close to zero at distances above $\sim$ 10 pc. This can be easily understood considering the fact that we set up the magnetic field with fluctuations down to $k$ = 128. This corresponds to a wavelength of 8 pc in a 1024 pc sized box thus implying a positive correlation only below this length in rough accordance with our results. For run M3\_B1\_SN where fallback occurs, the result does not change significantly at later times. Indeed the drop off is even somewhat steeper and $f(r)$ stays close to zero above $\sim$ 10 pc indicating a randomly distributed vector field above this scale.

For run M3\_B1\_PISN, however, we can see a clear evolution towards larger scales over time. The distance $r$ at which the curve intersects the x-axis increases over time and is about 165 pc for the last snapshot. This value of the intersection is about one order of magnitude larger than the value at $t$ = 0 in agreement with the shift of the peak in the magnetic field spectrum (right panel of Fig.~\ref{fig:spectrum}). Moreover, a wave length of 165 pc corresponds to wave-number of about 6 in a cube with a side length of 1024 pc, in good agreement with the position of the peak around $k$ = 5 in the corresponding spectrum.

\subsection{The influence of initial conditions and thermal properties}

We now focus on the influence of varying initial conditions and cooling ability on the properties of the magnetic field. Here, we only consider the end of each simulation but point out that the time evolution of the results is in general qualitatively similar to one of the two runs M3\_B1\_SN and M3\_B1\_PISN discussed before.

\subsubsection{Magnetic field strengths}
\label{sec:Bvary}

We first focus on the effect of changing the initial strength of the magnetic field. As listed in Table~\ref{tab:models}, we have performed further simulations with a 100 and 10$^4$ times stronger magnetic field focussing on the evolution of a PISN (run M3\_B2\_PISN and run M3\_B3\_PISN, respectively) in order to account for the uncertainty in the possible strength of the magnetic field in primordial star forming halos. For the case of a core collapse supernova we have repeated the simulation with a 100 times stronger magnetic field only (run M3\_B2\_SN). In this run the gas again starts to recollapse which is why qualitatively there are no significant differences in the spectrum compared to run M3\_B1\_SN and M3\_B1\_noSN. For this reason here we do not explicitly show the result of run M3\_B2\_SN. Furthermore, we consider the results of run M3\_B1\_5PISN, where we have assumed the simultaneous explosion of several PISN.
\begin{figure*}
 \includegraphics[width=0.48\linewidth]{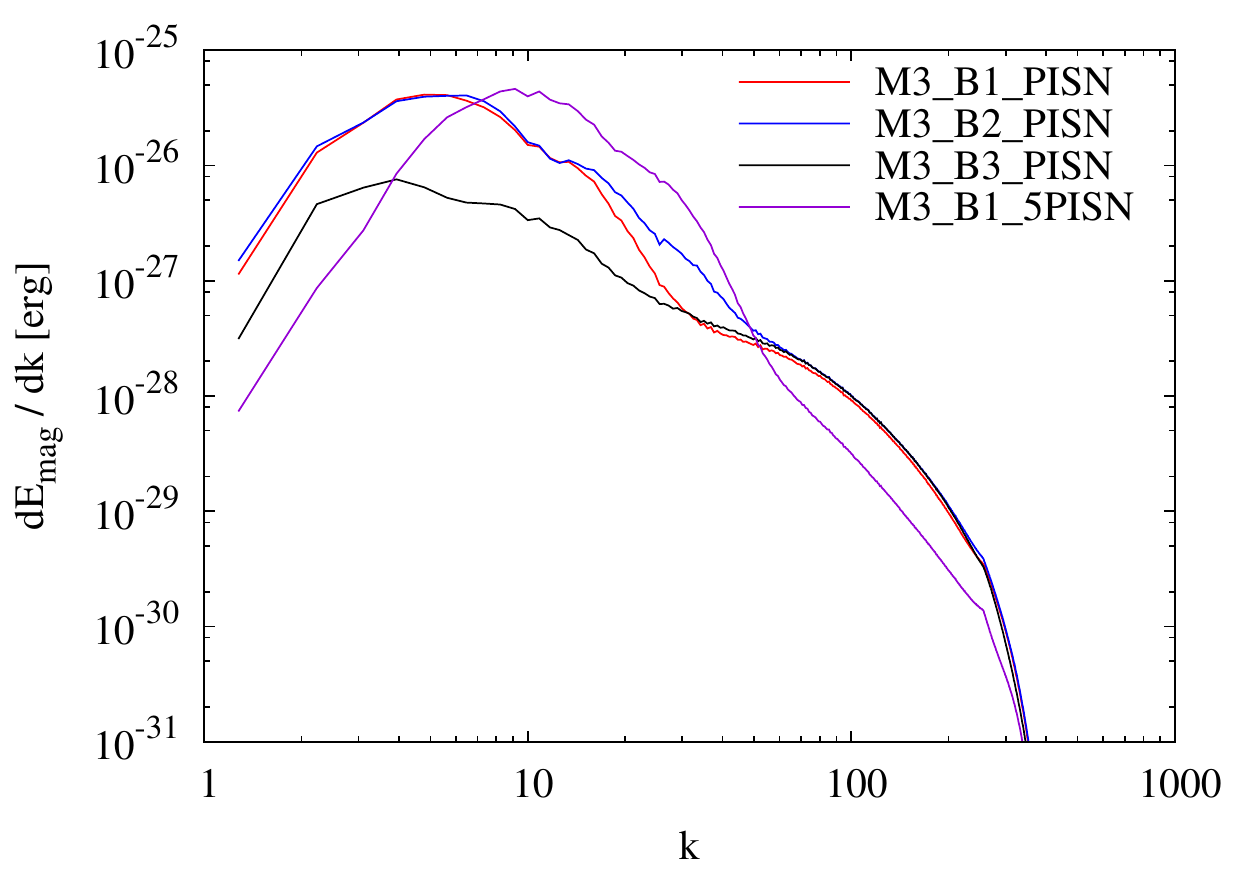}
 \includegraphics[width=0.48\linewidth]{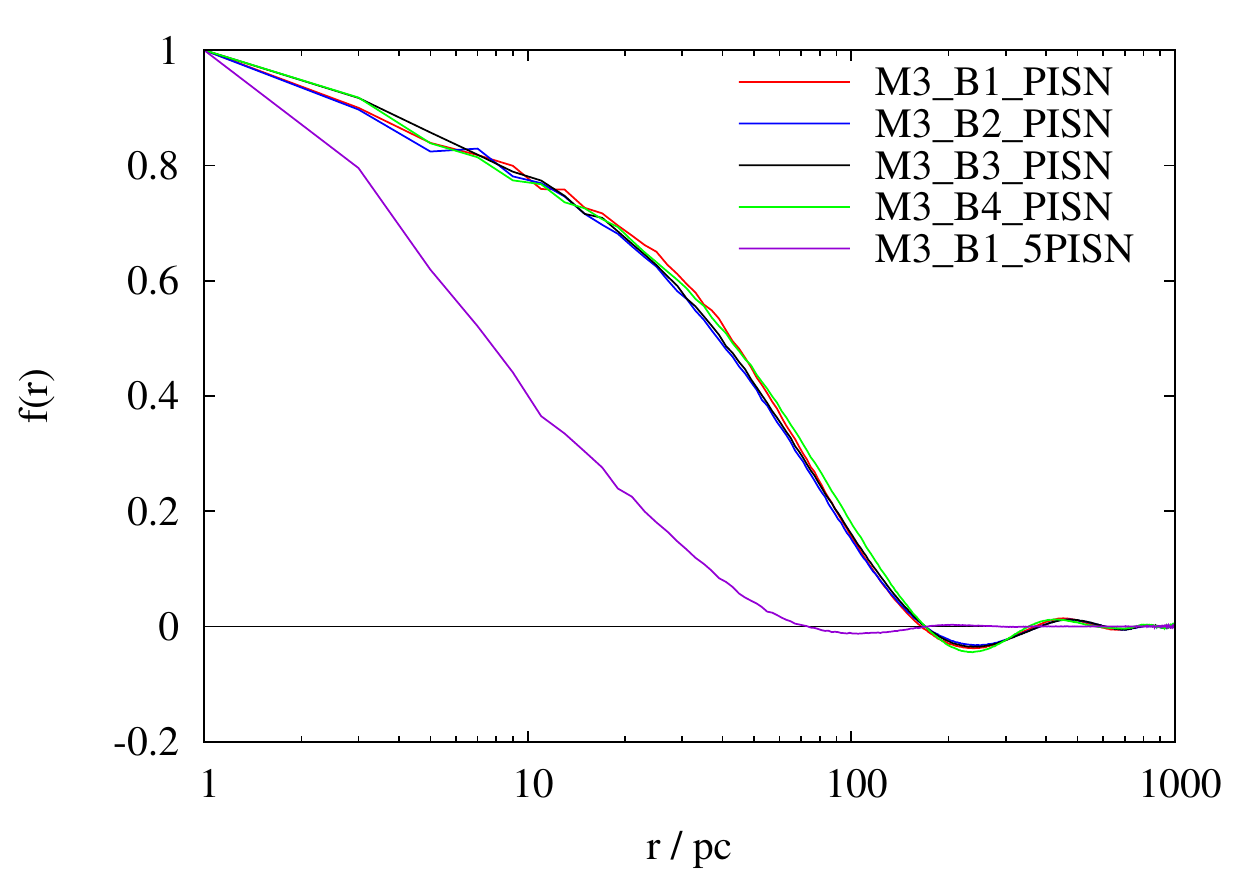}
\caption{Spectra (left) and autocorrelation function (right) of the magnetic field for the runs M3\_B1\_PISN, M3\_B2\_PISN, M3\_B3\_PISN, and M3\_B1\_5PISN at the end of each run. In addition, in the right panel the autocorrelation function of run M3\_B4\_PISN is shown. For comparative purposes, the spectra for run M3\_B2\_PISN and run M3\_B3\_PISN are scaled by a factor of 10$^{-4}$ and 10$^{-8}$, respectively.}
\label{fig:spectrumBvary}
\end{figure*}
The magnetic field spectra at the end of the aforementioned runs can be seen in the left panel of Fig.~\ref{fig:spectrumBvary}. We note that we have scaled the spectra of the runs M3\_B2\_PISN and M3\_B3\_PISN for comparative purposes by a factor of 10$^{-4}$ and 10$^{-8}$, respectively. As can be seen, the are some differences in the spectra, in particular at lower k. For run M3\_B3\_PISN the peak of the spectrum is less pronounced than for the runs with weaker magnetic fields. Nevertheless, in all cases with a single PISN the spectrum peaks around a $k$-value of 4 -- 5. For run M3\_B1\_5PISN, however, the spectrum peaks around 10 (see also discussion in Section~\ref{sec:nocool}).

\begin{figure}
 \includegraphics[width=\linewidth]{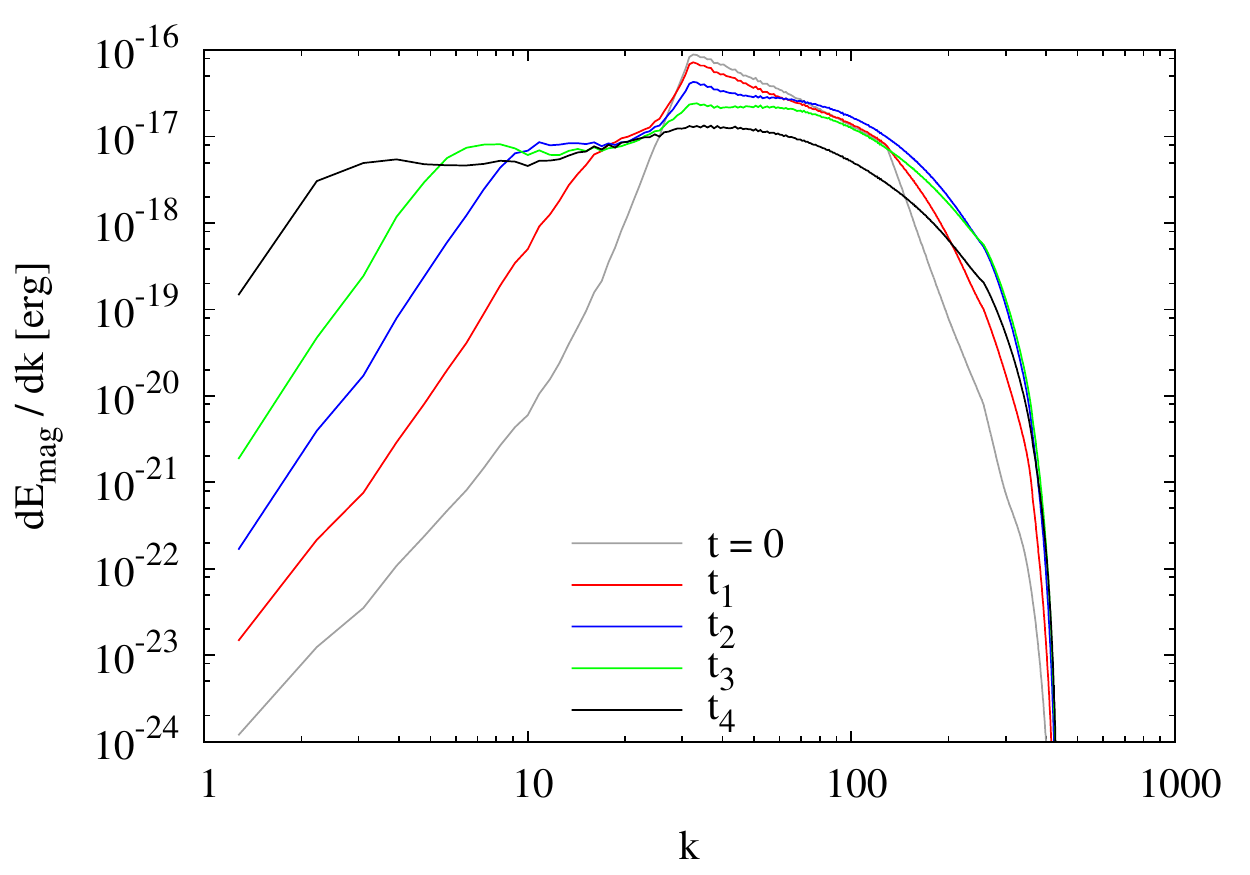}
\caption{Time evolution of the magnetic field spectrum for run M3\_B4\_PISN for the same times as in the right panel of Fig.~\ref{fig:spectrum}.}
\label{fig:D24_B0.1}
\end{figure}

Furthermore, we consider the spectrum of run M3\_B4\_PISN where the magnetic field is in equipartition with the turbulent, kinetic energy. Since for this run already the initial spectrum differs qualitatively from the other runs, we plot its time evolution separately in Fig.~\ref{fig:D24_B0.1} showing the spectra for the same times $t_1 - t_4$ as for run M3\_B1\_PISN (see Fig.~\ref{fig:spectrum} in Section~\ref{sec:spectra}). Despite the different shape of the spectrum, one basic results already observed in Section~\ref{sec:spectra} is reflected here: The peak of the spectrum is shifted by a factor of $\sim$ 10 to smaller $k$-values. But unlike for run M3\_B1\_PISN the peak suffers some initial damping, which is why it is not as pronounced as for run M3\_B1\_PISN.

We again point out that for the runs discussed here the magnetic energy stored on the peak scale remains relatively high over time. Only for the runs  M3\_B3\_PISN and M3\_B4\_PISN moderate losses occur over time whereas for the other runs the energy stored at the peak remains indeed almost unchanged. Since simultaneously on small scales the magnetic energy decreases over time, there is a net flow of energy from small to large scales resulting in the build-up of significantly strong, large-scale magnetic fields as discussed already before.

Next, we consider the effect of varying magnetic field strengths on the autocorrelation function $f(r)$ (right panel of Fig.~\ref{fig:spectrumBvary}) for the same runs as before. As can be seen, for the runs with a single PISN the autocorrelation functions at the end of each simulation are qualitatively and quantitatively very similar. In particular, the autocorrelation function of run M3\_B4\_PISN does not significantly differ from those of the other runs contrary to its spectrum (compare left panel of Fig.~\ref{fig:spectrumBvary} and Fig.~\ref{fig:D24_B0.1}). This is a direct consequence of the fact that for $f(r)$ we only account for the direction of the magnetic field and not its strength. Hence, the different initial scaling relations (see equations~\ref{eq:scaling} and~\ref{eq:scaling2}) are explicitly excluded. Moreover, in all four cases the magnetic field is dynamically not important compared to the supernova energy. We note, however, that in particular for the runs with a high initial magnetic field strength inside the remnant the magnetic pressure becomes comparable to the thermal pressure at late times. Further cooling of the gas would even increase the importance of the magnetic field for subsequent processes (see also Section~\ref{sec:lB}). For run M3\_B1\_5PISN we find $f(r)$ to decrease significantly faster in agreement with the spectrum shown in the left panel of Fig.~\ref{fig:spectrumBvary}. We attribute this to a reduced expansion time scale compared to the typical cooling time (see Section~\ref{sec:nocool}).

Finally, we try to estimate the field strength on large scales for the five runs discussed above. For this purpose we integrate the magnetic field spectra at the end of each run up to a $k$-value of 10 in order to obtain the magnetic energy stored on scales larger than 100 pc as already done in Fig.~\ref{fig:bfield} for run M3\_B1\_PISN. The corresponding magnetic field strengths range from about $10^{-3}$ to $10^1$ nG depending on the initial field strength (see Table~\ref{tab:results}). It can be seen that the ratio between the different large-scale field strengths roughly corresponds to the initial ratio of the field strengths at 80 pc (see column 5 of Table~\ref{tab:models}), only for run M3\_B4\_PISN the ratio is somewhat lower. Furthermore, we find that at the end of each run the magnetic energy at $k$ $<$ 10 has experienced at \textit{relative} increase by about 3 orders of magnitude compared to the initial energy in this range. For the larger-scale version of run M3\_B1\_PISN (box size of 2 kpc) we measure the increase of magnetic energy above 200 pc (again $k$ $<$ 10). Here, a relative increase of energy by even more than 4 orders of magnitude occurs. Moreover, for all runs considered the magnetic energy on large scales (k $<$ 10) constantly increases over time --  or at least remains constant but never decreases, whereas the energy on smaller scales decreases (compare Fig.~\ref{fig:bfield}). This confirms our assumption made at the end of Section~\ref{sec:spectra} that the large-scale magnetic field builds up at the cost of the small-scale fluctuations.
\begin{table}
 \caption{Coherence length $l_\rmn{B}$ of the magnetic field at the end of those simulations where the supernova remnant keeps expanding. In column 3 and 4 the magnetic field strength on large scales (k $<$ 10, see text) at the end and the beginning of the corresponding run is shown.}
 \label{tab:results}
\centering
 \begin{tabular}{@{}lccc}
  \hline
 Run & l$_\rmn{B}$ & $B$(k$<$10)$_\rmn{end}$ & $B$(k$<$10)$_\rmn{init}$ \\
     & [pc] & [nG] & [nG] \\
  \hline
 M1\_B1\_SN & 182 & $2.05 \cdot 10^{-3}$ & $6.53 \cdot 10^{-5}$ \\
 M1\_B1\_PISN & 112 & $1.97 \cdot 10^{-3}$ & $6.53 \cdot 10^{-5}$ \\
\hline
 M2\_B1\_PISN & 94 & $2.08 \cdot 10^{-3}$ & $6.53 \cdot 10^{-5}$ \\
\hline
 M3\_B1\_PISN & 214 & $2.43 \cdot 10^{-3}$ & $6.53 \cdot 10^{-5}$  \\
 M3\_B1\_5PISN & 112 & $2.16 \cdot 10^{-3}$ & $6.53 \cdot 10^{-5}$  \\
\hline
 M3\_B2\_PISN & 158 & $2.48 \cdot 10^{-1}$ & $6.53 \cdot 10^{-3}$  \\
 M3\_B3\_PISN & 260 & $1.03 \cdot 10^{1}$ & $6.53 \cdot 10^{-1}$  \\
 M3\_B4\_PISN & 260 & $3.08 \cdot 10^{1}$ & $4.71 \cdot 10^{-1}$ \\
\hline
 \end{tabular}
\end{table}

\subsubsection{Masses of the halo}

\begin{figure*}
 \includegraphics[width=0.48\linewidth]{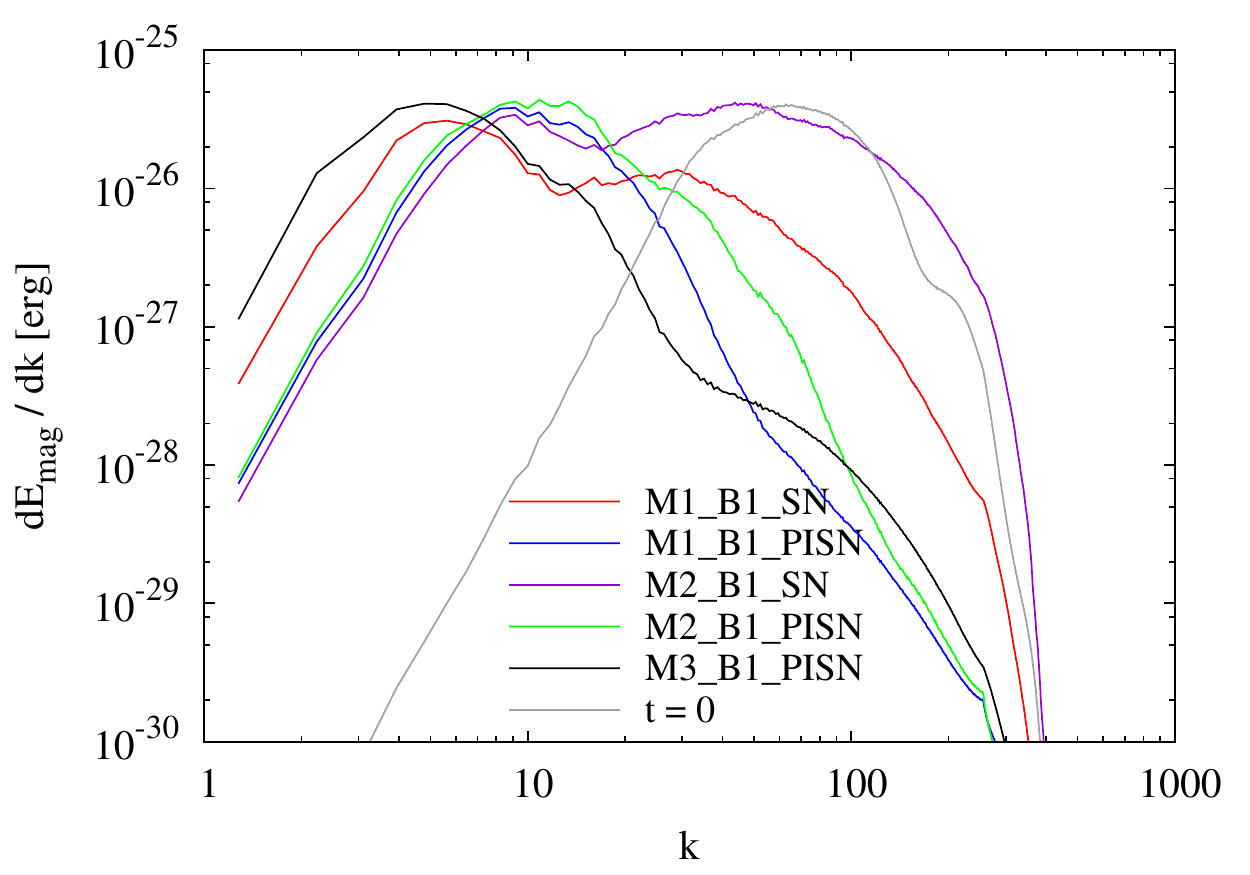}
 \includegraphics[width=0.48\linewidth]{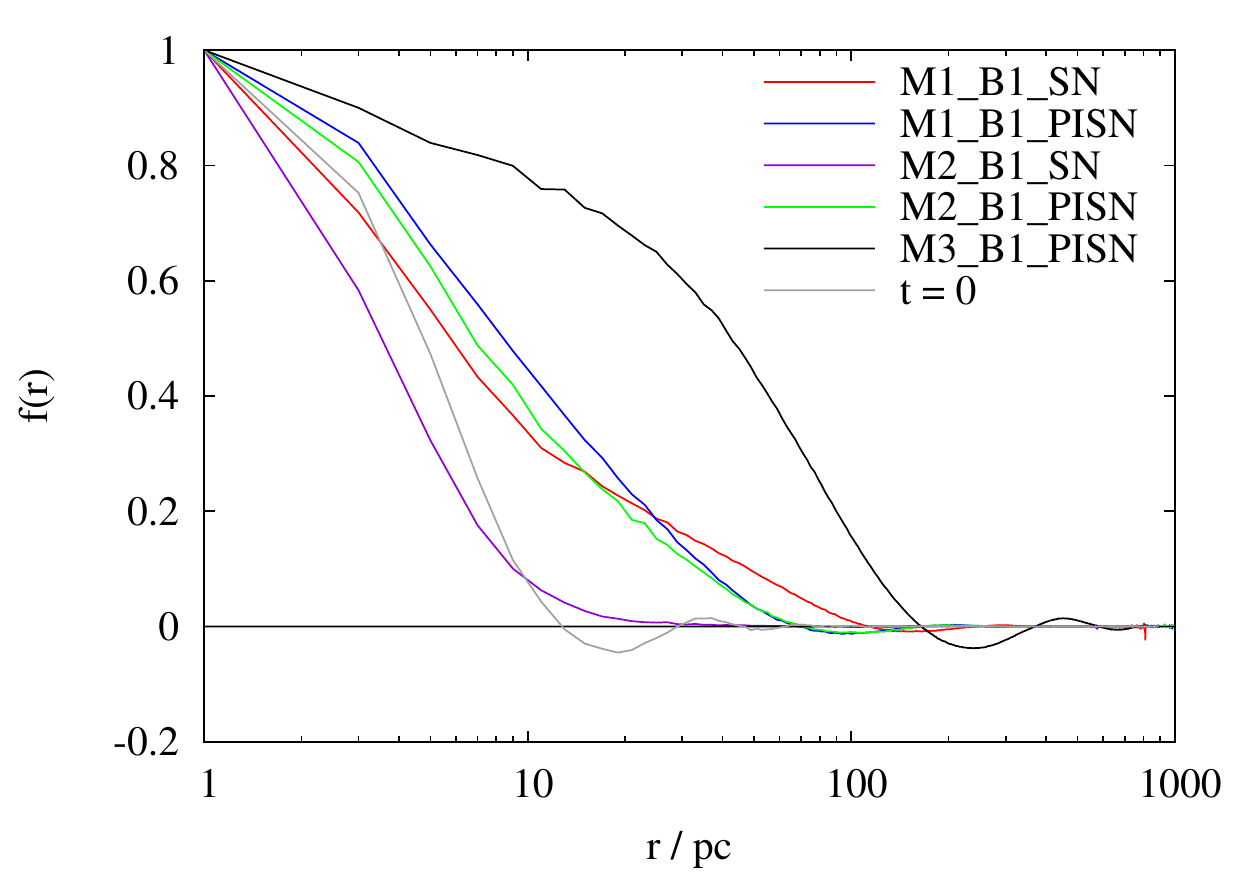}
\caption{Left: Magnetic field spectra for the runs M1\_B1\_SN, M1\_B1\_PISN, M2\_B1\_SN, M2\_B1\_PISN, and M3\_B1\_PISN at the end of each simulation. Right: Autocorrelation function for the same runs as in the left panel. For comparative purposes in both panels the result at $t$ = 0 is shown (grey line).}
\label{fig:Mvary}
\end{figure*}

Next, we investigate the effect of the halo mass. Since we have seen in Section~\ref{sec:Bvary} that changing the magnetic field strength over a wide range does not significantly alter the results, we concentrate on a single value for the magnetic field. In the left panel of Fig.~\ref{fig:Mvary} we plot the spectra of the runs M1\_B1\_SN, M1\_B1\_PISN, M2\_B1\_SN, M2\_B1\_PISN, and M3\_B1\_PISN at the end of each simulation\footnote{The runs M1\_B1\_SN and M2\_B1\_SN were followed for 10 Myr and the other three runs until the shock front reached the boundary of the simulation domain.}. We point out that for run M3\_B1\_SN showing recollapse the spectrum as well as the autocorrelation function are not shown.
For the runs M1\_B1\_SN, M1\_B1\_PISN, M2\_B1\_PISN, and M3\_B1\_PISN, where the remnant is able to disrupt the halo and escape from the gravitational potential well, the spectra peak at low $k$-values of the order of 5 -- 10 in agreement with the results of the previous sections. For run M2\_B1\_SN is less clear. Here the supernova remnant has already started to recollapse in the central regions, whereas in the outer regions the remnant keeps expanding even after 10 Myr. Hence, we see a bimodal spectrum with one peak coming from the ongoing expansion and the peak at small scales from the beginning recollapse.

The autocorrelation functions for the runs M1\_B1\_SN, M1\_B1\_PISN, M2\_B1\_SN, M2\_B1\_PISN, and M3\_B1\_PISN are shown in the right panel of Fig.~\ref{fig:Mvary}. For the all runs except run M2\_B1\_SN the correlation length increases by a factor $\sim$ 10 reaching values of the order of 100 -- 200 pc. However, it can be seen that for the runs with the lower halo masses the correlation seems to be less strong on larger scales compared to run M3\_B1\_PISN (black line) as indicated by the steeper decline in $f(r)$. This nicely agrees with the finding that the spectrum of run M3\_B1\_PISN is the one with the most magnetic energy on large scales, i.e. that it peaks at the smallest $k$-value (see left panel of Fig.~\ref{fig:Mvary}). For run M2\_B1\_SN (violet line) the decline is even stronger in agreement with the double-peaked spectrum.

\subsubsection{Cooling ability}
\label{sec:nocool}

Recently, \citet{Gnedin12} have shown that the cooling ability of primordial gas can be significantly reduced in regions close to a source emitting large amounts of ionizing radiation. For this reason, we have simulated an extreme case in which the cooling is completely turned off, i.e. the gas behaves purely adiabatic.

\begin{figure}
 \includegraphics[width=\linewidth]{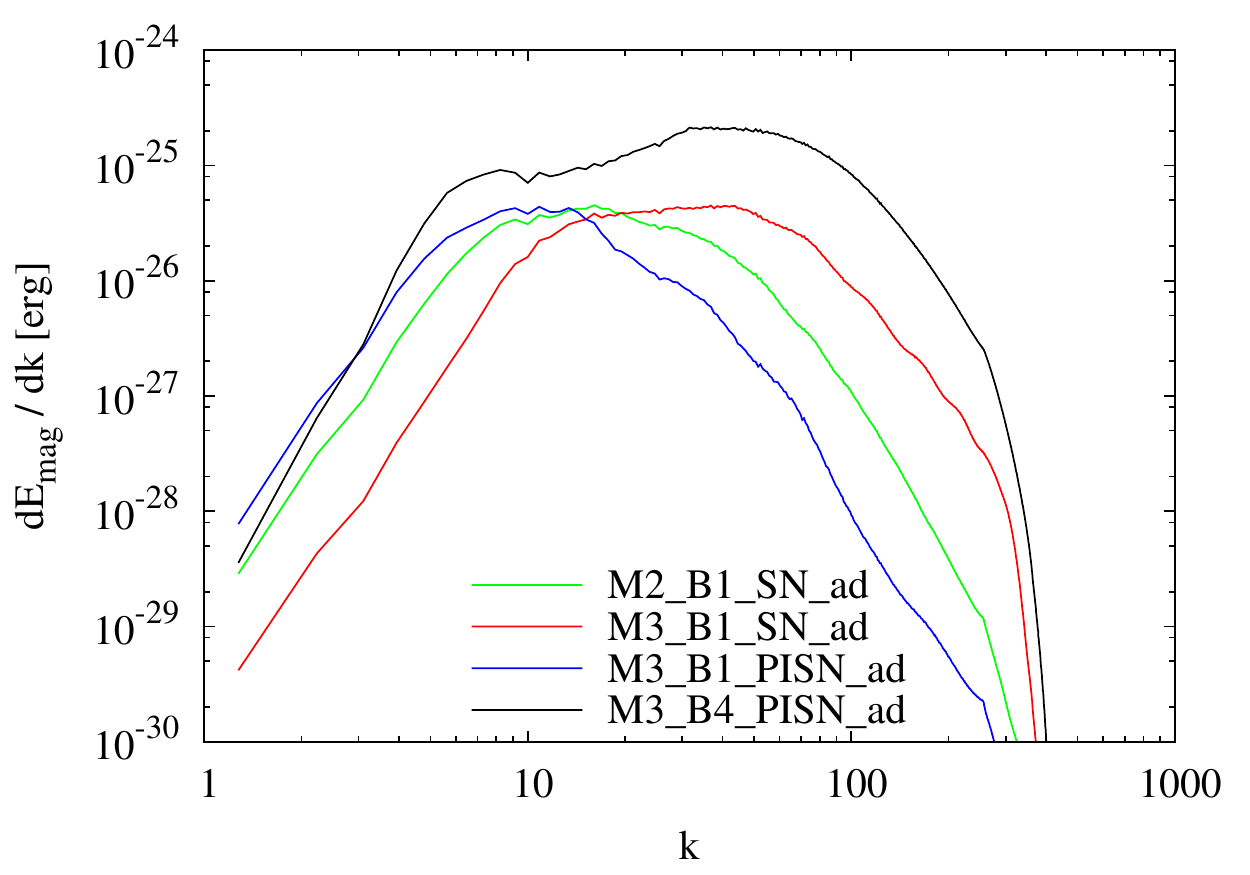}
\caption{Magnetic field spectra of the runs M2\_B1\_SN\_ad, M3\_B1\_SN\_ad, M3\_B1\_PISN\_ad, and M3\_B4\_PISN\_ad at the end of each simulation. For comparative purposes the spectrum of run M3\_B4\_PISN\_ad is scaled by a factor of $10^{-8}$.}
\label{fig:cool}
\end{figure}

We first point out that in run M2\_B1\_SN\_ad and M3\_B1\_SN\_ad the core collapse supernova manages to disrupt the halo in contrast to the runs M2\_B1\_SN and M3\_B1\_SN with cooling. This can be understood on the basis of a simple energy argument: The gravitational binding energies of the baryonic gas for the runs with a total halo mass of $7.5 \cdot 10^5$ and $4.2 \cdot 10^6$ M$_{\sun}$ are $2.7 \cdot 10^{49}$ and $5.7 \cdot 10^{50}$ erg, respectively. Since both values are smaller than the explosion energy of the supernova (E$_{\rmn{SN}} = 10^{51}$ erg) and since no energy is lost by cooling processes, the supernova can disrupt the halo. We now show the magnetic field spectra for the runs M2\_B1\_SN\_ad, M3\_B1\_SN\_ad, M3\_B1\_PISN\_ad, and M3\_B4\_PISN\_ad at the end of each run in Fig.~\ref{fig:cool}. 
The autocorrelation function for those runs (not shown here) agree well with the results of the spectrum analysis. As can be seen in Fig.~\ref{fig:cool}, the spectra of the runs without cooling in general peak a somewhat higher $k$-values than the corresponding runs with cooling. The higher value of k$_\rmn{peak}$ is similar to the findings in the runs with a low halo mass (the runs M1\_B1\_SN, M1\_B1\_PISN, and M2\_B1\_PISN) as well to the findings of run M3\_B1\_5PISN, which all peak at somewhat smaller $k$-values than run M3\_B1\_PISN. We attribute this behaviour to the same physical origin: For the lower-mass halos the cooling ability of the gas is reduced due to the overall lower densities (see equation~\ref{eq:cooling}) and the gas requires a longer time to cool. Similarly, for run M3\_B1\_5PISN a longer cooling time is required due to the higher explosion energy. In consequence, in both cases the cooling time scale is prolonged when compared to the typical expansion timescale of the supernova.

\begin{figure}
\centering
 \includegraphics[width=0.75\linewidth]{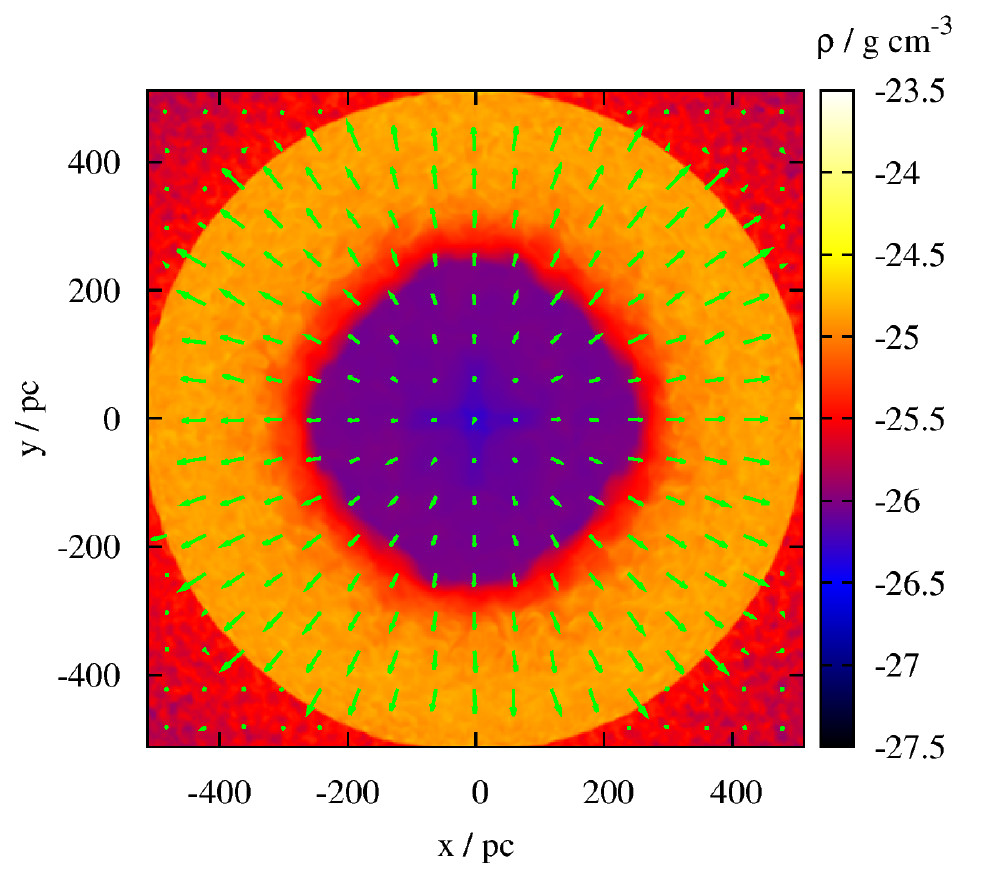} \\
 \includegraphics[width=0.75\linewidth]{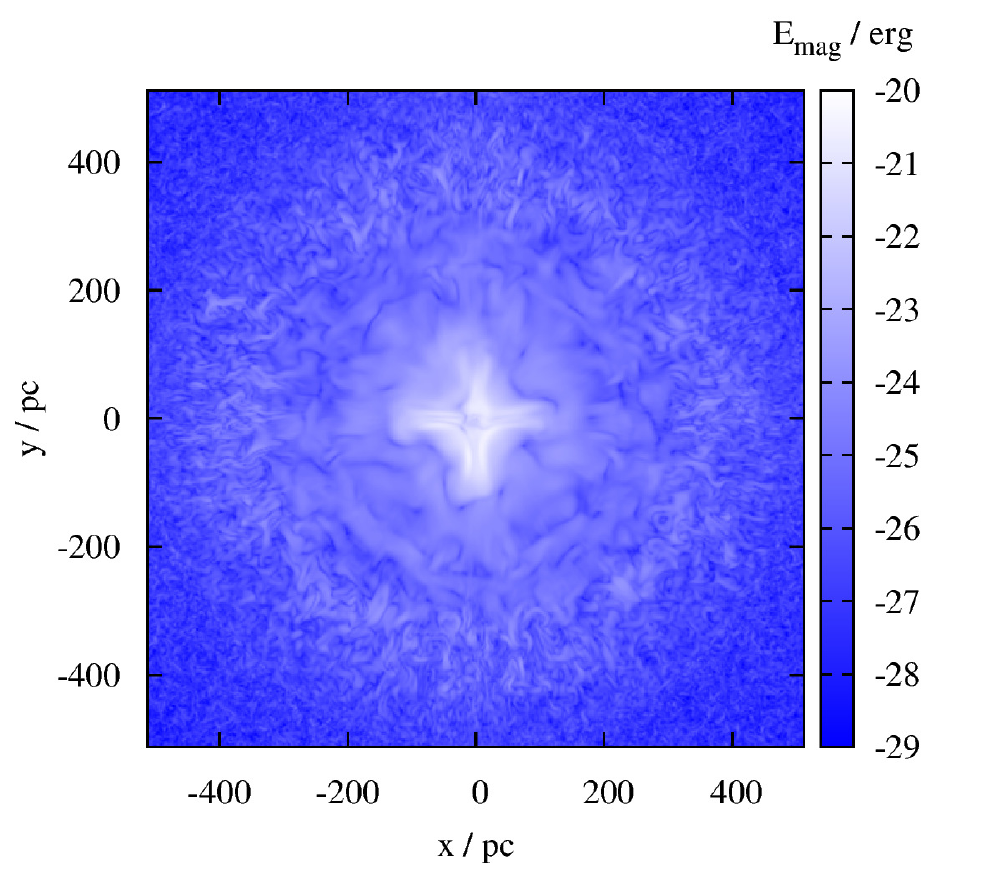}
\caption{Slice through the center of run M3\_B1\_PISN\_ad showing the density (top) and magnetic energy (bottom) at the end of the simulation. The high-density layer behind the shock front is much thicker than in the corresponding run with cooling and contains strong, small-scale magnetic field fluctuations.}
\label{fig:nocool}
\end{figure}

For this reason, the evolution of the gas is more similar to an adiabatic evolution like in the runs in which cooling is turned off completely. Hence, for these runs as well as the runs without cooling a rather thick shell develops in contrast to the thin shock fronts seen in the other runs. This is demonstrated exemplarily for run M3\_B1\_PISN\_ad in Fig.~\ref{fig:nocool} where the density and magnetic energy at a slice through the center are shown. The high-density region behind the shock front is much thicker than for the case with cooling (see Fig.~\ref{fig:slicesSN53}) and contains significant, small-scale magnetic field fluctuations not observed in run M3\_B1\_PISN. These small-scale fluctuations cause the spectra to peak at a higher values of k. Similar results are also found for the runs M1\_B1\_SN, M1\_B1\_PISN, M2\_B1\_PISN, and M3\_B1\_5PISN, hence for all the runs in which the cooling time is increased relative to dynamical timescale.

\section{Discussion}
\label{sec:discussion}

\subsection{Halo disruption}
\label{sec:disruption}

\begin{figure}
 \includegraphics[width=\linewidth]{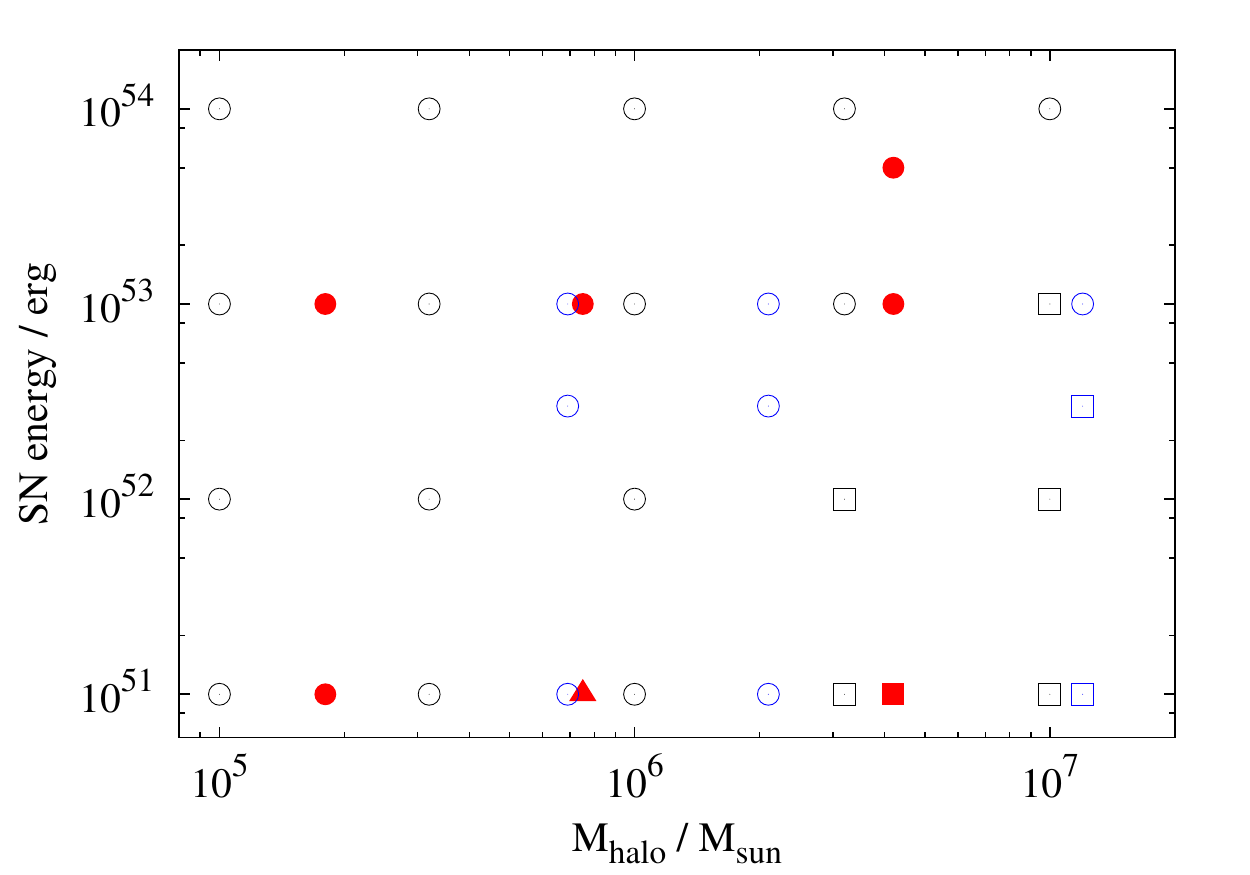}
\caption{Final fate of the supernova remnant in dependence of the halo mass and explosion energy. Circles denote the disruption of the halo by the supernova, squares denote fallback. Red, filled points mark the result of this work, open black and blue points the results of~\citet{Kitayama05} and~\citet{Whalen08}, respectively. For run M2\_B1\_SN (red triangle) the situation is unclear (see text).}
\label{fig:phase}
\end{figure}

We consider the final fate of the supernova remnant in dependence of the halo mass and the injection energy and compare the results with those from other numerical studies. To remind the reader we find that a core collapse supernova ($E_\rmn{SN} = 10^{51}$ erg) cannot disperse halos as massive as $4 \cdot 10^6$ M$_{\sun}$ or above. Also for the halo with $7.5 \cdot 10^5$ M$_{\sun}$ (run M2\_B1\_SN) we see signs of recollapse in the central region although the outer regions keep expanding. Hence, this mass seems to be roughly the borderline between expansion and recollapse. We note, however, that we would have to include a more detailed treatment of the thermodynamics as well as radiation transfer in order to make a definite statement about the final state of the supernova remnant. In general, however, our findings are in agreement with recent works of~\citet{Bromm03} and~\citet{Ritter12} modelling the evolution of a core collapse supernova in a halo of $\sim$ 10$^6$ M$_{\sun}$. Two further parameter studies on the final fate of primordial supernova without magnetic fields were performed by~\citet{Kitayama05} and~\citet{Whalen08}. We combine our results with the results of these authors in Fig.~\ref{fig:phase} showing the final fate of the supernova in dependence of the halo mass and the explosion energy.
As can be seen, our results agree reasonably well with the results of~\citet{Kitayama05} and~\citet{Whalen08}. The combination of the different results suggest that supernovae with an explosion energy of $10^{51}$ erg are able to disrupt halos with masses up to $\sim$ $10^6$ M$_{\sun}$. For PISN, our results agree with that of \citet{Whalen08} and \citet{Kitayama05}. The apparent contradiction between the result of \citet{Whalen08} and \citet{Kitayama05} might be due to the fact that the latter authors initialized their explosions with purely thermal energy, which might be radiated away too quickly \citep{Whalen08}. A fraction of the explosion energy should rather be inserted as kinetic energy as it was done in this work. Hence, this would imply a mass limit of the of about $10^7$ M$_{\sun}$, up to which halos can be dispersed by a PISN.

Our findings that the core collapse supernova is not able to disrupt the halo with a mass of $4.2 \cdot 10^6$ M$_{\sun}$ -- despite the explosion energy being somewhat larger than the gravitational binding energy of the baryons (see Section~\ref{sec:nocool}) -- agrees with the findings of \citet{Kitayama05}: For massive halos explosion energies significantly larger than the gravitational binding energy are required to disrupt the halo. This is due to the efficient radiative cooling at high densities removing energy before it can contribute to the disruption of the halo.

Overall, we find that -- despite the strong simplifications made to model our initial conditions and the highly idealized treatment of the thermodynamical gas properties -- our results concerning the final fate of the halo are in reasonable agreement with simulations including a much more sophisticated treatment of the physical processes involved \citep{Bromm03,Kitayama05,Whalen08,Ritter12}.

\subsection{Coherence length of magnetic fields}
\label{sec:lB}

\begin{figure}
 \includegraphics[width=\linewidth]{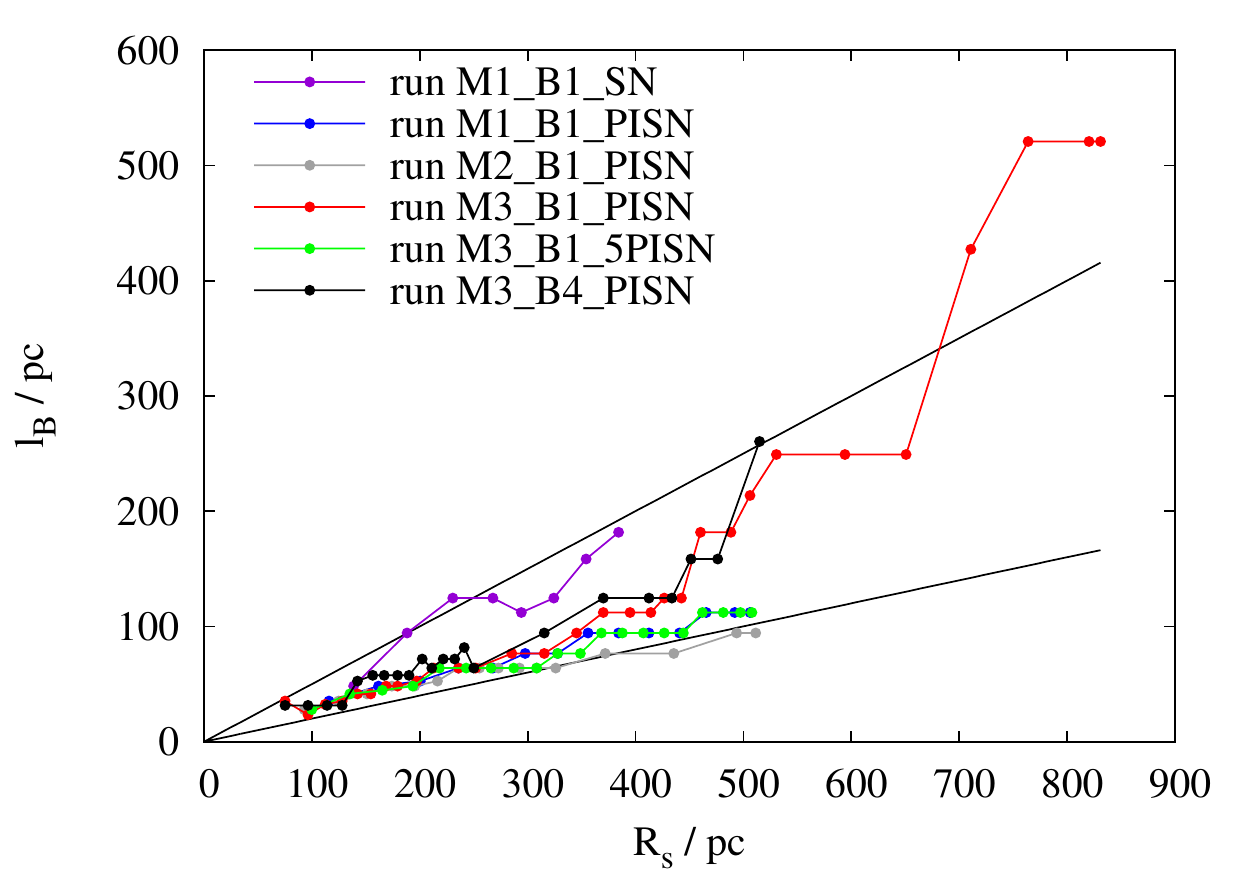}
\caption{Coherence length $l_\rmn{B}$ of the magnetic field plotted against the radius of the supernova shell for the runs M1\_B1\_SN, M1\_B1\_PISN, M3\_B1\_PISN, M3\_B1\_5PISN, and M3\_B4\_PISN. For all cases the coherence length reveals an approximately linear scaling with $R_\rmn{s}$. To guide the readers eye, we show a linear relation of $l_\rmn{B} = 0.2 \cdot R_\rmn{s}$ and $l_\rmn{B} = 0.5 \cdot R_\rmn{s}$, respectively (black lines).}
\label{fig:lB}
\end{figure}

From the magnetic field spectra discussed before we can infer the time evolution of the coherence length $l_\rmn{B}$ of the magnetic field over time. For the sake of simplicity we define the coherence length as
\begin{equation}
 l_\rmn{B} = \frac{1024}{k_\rmn{peak}} \, \rmn{pc} \, ,
\label{eq:lB}
\end{equation}
where $k_\rmn{peak}$ is the (dimensionless) $k$-value of the peak of the spectra from our simulations. Since the expansion of the supernova remnant is not linear in time (see Fig.~\ref{fig:Rshell}) and since we are more interested in how the coherence length relates to the radius of the supernova shell $R_\rmn{s}$, in Fig.~\ref{fig:lB} we plot $l_\rmn{B}$ against the $R_\rmn{s}$ for several selected simulations\footnote{For run M3\_B1\_PISN we have taken the result of the larger-scale run with a box size of 2048 pc followed over 10 Myr.}.
As can be seen, the coherence length $l_\rmn{B}$ keeps on increasing with increasing shell radius for all cases considered. The two black lines show an assumed linear scaling of the coherence length $l_\rmn{B}$ as $0.2 \cdot R_\rmn{s}$ and $0.5 \cdot R_\rmn{s}$. It can be seen that the coherence length of the magnetic field inside a supernova bubble reaches values of the order of 100 -- 300 pc (see also Table~\ref{tab:results}) and thus of the order of 20\% -- 50\% of the radius of the shell. For late times in run M3\_B1\_PISN the coherence length even reaches values of up to $\sim$ 60\% of the shell radius. We note that in order to improve the readability of the plot we do not show the results of the runs M3\_B2\_PISN and M3\_B3\_PISN, in particular since qualitatively they do not differ too much from run M3\_B1\_PISN.

\citet{Greif07} simulate the evolution of a PISN from redshift 20 to 12, i.e. over 200 Myr until the shock front stalls. By this time the shock front has reached a distance of about 2.5 -- 3 kpc from the progenitor star. Hence, under the assumption that the approximately linear relation between the coherence length of the magnetic field and the supernova shell radius found in our work holds over the \textit{entire} evolution of the remnant, this results in ordered magnetic fields on scales as large as $\sim$ 1.5 kpc.

Moreover, as pointed out in the Sections~\ref{sec:spectra} and~\ref{sec:Bvary}, the amount of magnetic energy on large scales remains approximately constant or even increase over time at the cost of the small-scale fluctuations. Hence, for a region affected by a supernova one could expect a significant magnetic field to be present on large scales even at late times. Depending on the efficiency of the small-scale dynamo during the lifecycle of the Pop III stars, we find a large-scale ($>$ 100 pc) magnetic field component with a strength of up to 10$^{-8}$ G, at least for an initially saturated field.

\subsection{Effects during the recollapse of halos}

Such a large-scale magnetic field as discussed before could have a significant effect on the possible recollapse of the bubble interior, on the ensuing potential formation of supermassive black holes \citep{Whalen13a,Whalen13b,Johnson13}, and on the overall fragmentation properties of the gas and the ability to form protostellar disks \citep{Machida13}. The latter authors showed that for primordial minihalos with a well-ordered magnetic field and a mass-to-flux ratio \citep{Mouschovias76} below $\sim 10^4$ fragmentation is largely reduced. In order to make some predictions about the expected fragmentation properties during a potential second collapse in our case, we approximate the mass-to-flux-ratio by
\begin{equation}
 \mu = \left( \frac{\Sigma}{B} \right) / \left( \frac{\Sigma}{B} \right)_{\rmn{crit}} = \left( \frac{\rho \cdot l_\rmn{B}}{B} \right) / \left( \frac{\Sigma}{B} \right )_\rmn{crit} \, ,
\end{equation}
with the critical mass-to-flux ratio $(\Sigma / B_\rmn{crit}) = 0.13/\sqrt{G}$ and the gravitational constant G \citep{Mouschovias76}, the coherence length $l_\rmn{B}$ of the magnetic field  ranging from 100 -- 500 pc (see Fig.~\ref{fig:lB}), $\rho \sim 10^{-27}$~g~cm$^{-3}$ (see bottom left panel of Fig.~\ref{fig:slicesSN53}), and the field strength $B$ from $\sim 2 \cdot 10^{-12}$ G -- $ 3 \cdot 10^{-8}$ G (see Table~\ref{tab:results}). With these values we obtain mass-to-flux ratios in the range of 0.02 -- 1500. Hence, under the reasonable assumption that during a subsequent recollapse the mass-to-flux ratio is more or less preserved and based on the results of \citet{Machida13}, we would expect a relatively low degree of fragmentation in such a minihalo undergoing a second round of star formation.

\subsection{Effects on first galaxy formation and the IGM}

As shown here, in regions with a high halo density the occurrence of several supernovae might lead to the generation of a volume filling, relatively well-ordered magnetic field of significant strength. In particular, the bottom-up formation of the first galaxies is believed to occur in regions with a significantly higher halo density than predicted under the assumption of a homogeneous distribution. \citet{Gao10} find a mean separation between potentially star forming halos of the order of $0.5 - 1$ kpc. Since this separation is of the order of or even somewhat smaller than the typical coherence length of the magnetic field inferred in this work ($\leq 1.5$ kpc), one could indeed expect a volume filling, large-scale magnetic field in regions where the first galaxies have formed.

As shown by~\citet{Heger02}, only stars with masses between $\sim$ 10 -- 50 M$_{\sun}$ and between 140 -- 260 M$_{\sun}$ will end their lives with a core collapse supernova or a PISN, respectively. In this context, we point out that despite the expected fragmentation in primordial halos and the formation some low-mass Pop III stars \citep{Clark11,Greif12,Stacy13}, recent simulations have shown that star-forming halos should host Pop III stars with typical masses ranging from 10 M$_{\sun}$ up to about 50 M$_{\sun}$ limited by radiative feedback \citep{Hosokawa11,Latif13c,Susa13} but also a good fraction with masses in excess of 100 M$_{\sun}$ \citep{Hirano14}. Hence, one can expect that masses of Pop III stars indeed lie in the range suitable for a core collapse supernova or a PISN and that each star forming halo is affected by one or more supernovae. Moreover, around a redshift $z = 20$ the Press-Schechter formalism~\citep{Press74} predicts only few halos to have virial masses significantly above $10^6$ M$_{\sun}$. Therefore it is reasonable to assume that -- despite the fact that a core collapse supernova can only disrupt halos up to a few 10$^6$ M$_{\sun}$ as discussed in Section~\ref{sec:disruption} -- most of the times a halo is forming Pop III stars, it will be disrupted by a subsequent supernova resulting in the build-up of a large-scale magnetic field.

Hence, the volume affected by supernovae during the assembly of first galaxies seems to depend crucially on the exact number of star forming halos. \citet{Greif08} and \citet{Wise08} simulating the formation of a first galaxy including stellar feedback find that of the order of 10 halos form Pop III stars prior to the formation of the galaxy. The simulations presented here thus suggest that a significant fraction of the volume of a region forming a protogalaxy might indeed be pervaded by a large-scale magnetic field of considerable strength. Such a magnetic field configuration would possibly also affect the formation of the first galaxies and in particular their early evolution: It is usually assumed that the initial magnetic field in protogalaxies is rather small-scaled. Subsequently, a kpc-scale magnetic field of considerable strength builds up within about 10$^8$ yr or less due to the small-scale dynamo \citep{Arshakian09,Schober13}. This field serves as a seed for the mean-field dynamo responsible for the large-scale, well-ordered magnetic field observed in present day galaxies~\citep[e.g.][]{Beck94}. Hence, a large-scale magnetic field from the very beginning as suggested here could possibly allow the mean-field dynamo to act at an even earlier phase thus potentially affecting the later evolution of the galaxy.

Finally, we note that the emergence of highly energetic pair instability supernovae in lower-mass halos could also contribute to the magnetization of the intergalactic medium (IGM). \citet{Kronberg99} showed that supernova-driven winds in dwarf galaxies can magnetize the IGM by a substantial degree starting around z = 10. We tentatively suggest that -- depending on the frequencies of pair instabilities supernovae in primordial halos -- the magnetization of the IGM could have started already around z = 20. A magnetization on kpc scale with field strengths of 10$^{-12}$ -- 10$^{-8}$~G found in this work (see Table~\ref{tab:results}) would be in good agreement with recent Faraday rotation measurements of \citet{Neronov13} as well as with FERMI observations, revealing lower bounds of the magnetic field on kpc scales of the order of $10^{-14}$ G \citep[e.g.][see their figure 2.]{Neronov10}.

\subsection{Modelling uncertainties}
\label{sec:caveats}

We emphasize that in order to properly analyse the properties of the magnetic field -- the main focus of this work --  we had to use a high and homogeneous spatial resolution of 512$^3$ grid points in our simulations. On the other hand, simulations making use of the adaptive mesh refinement algorithm and refining e.g. for the gas density, usually reveal a well resolved region in the center of the halo but rather poorly resolved outer parts. Such a grid structure, however, would be highly disadvantageous for analysing the magnetic field structure by means of a Fourier transformation as done in Section~\ref{sec:spectra}. Moreover, due to the large number of runs required to explore the parameter space -- consisting of the halo mass, the magnetic field strength and the explosion energy -- a fully self-consistent treatment of all relevant physical processes like self gravity, chemistry, cooling and radiation transport is currently not possible. For this reason we have not applied self gravity in the simulations but used a time-independent gravitational potential.

Furthermore, we modelled the thermodynamical behaviour of the gas by means of a tabulated cooling curve~\citep{Sutherland93} making use of a subcycling scheme for updating the thermal energy. It was argued that such a subcycling scheme might be too inaccurate and that the global minimum of the Courant timestep and the cooling timestep should rather be used to update the hydrodynamics \citep[e.g.][in particular their Section 2.4]{Whalen06}. However, since we are mainly interested in how the supernova globally reorganises the magnetic field structure, for our purposes we consider our approach as sufficient. We furthermore note that due to the lack of the Compton cooling process in our cooling routine, we somewhat underestimate the real cooling efficiency. This, in turn, probably causes the final radii of our remnants (Fig.~\ref{fig:lB}) to be somewhat too large. However, since our results concerning the radial expansion agree reasonably well with that of more sophisticated simulations \citep[see, e.g. Fig. 7 of][and Fig. 10 of \citealt{Whalen08}]{Greif07}, the actual effect of the reduced cooling efficiency on the remnant size is hard to assess. Finally, we note that the cooling efficiency in our work as well as the works mentioned before might generally be too low since cosmic ray acceleration and subsequent emission due to second order Fermi processes in the shock front have not been considered \citep{Fermi49}. This effect could again result in somewhat smaller final radii of the supernova remnants.

We suppose that a more sophisticated inclusion of all effects mentioned before would cause at least 10 -- most likely even more -- times higher computational costs. However, since we made a great effort to generate proper initial conditions in particular for the magnetic field and since we find a good agreement with more sophisticated simulations~\citep[e.g.][]{Bromm03,Kitayama05,Whalen08,Ritter12}, we consider our simplified approach as justified.

We note that setting the initial density in the supernova remnant and the HII region to a constant value (see Fig.~\ref{fig:IC}), is a rather crude approximation. However, we would like to point out that due to the lack of spatial resolution we were not able to self-consistently model the earliest evolutionary phase of the supernova. Moreover, at later times density and velocity profiles in our simulations are in rough agreement with other simulations \citep[e.g.][]{Kitayama05,Whalen08}. For this reason, and since we are mainly interested in the late stages of the supernova evolution, we consider the approximations made in our initial setup as sufficient. However, as shown by \citet{Whalen08}, a reverse shock, occurring once the supernova shock front hits the edge of the HII region, might trap a fraction of the magnetic energy in the center thus possibly reducing the strength of the large-scale magnetic field, a process we are clearly lacking in our setup. Hence, in combination with the possible overestimation of the final size of the remnants (due to a simplified cooling description as discussed above) we suggest that the values concerning the coherence length and strength of the large-scale magnetic field given in this work (Table~\ref{tab:results}) should rather be considered as upper limits.

The simulations presented here can also serve as a valuable guide for future and more sophisticated simulations including a better treatment of processes like gravity, chemistry, and/or radiation transfer. In particular, it is planned to repeat some of the selected simulations with self gravity and a proper treatment of the chemistry making use of a new primordial chemistry module implemented in FLASH~\citep{Grassi13}\footnote{http://kromepackage.org/}. By doing so we can test to what extent the magnetic field properties depend on the proper treatment of these processes. Moreover, as a next future step it is also planned to take the initial conditions directly from cosmological simulations~\citep[e.g.][]{Latif13d,Latif13a,Latif13b}.

\section{Conclusions}
\label{sec:conclusion}

We have presented simulations of core collapse supernovae and pair instability supernovae going off in the center of magnetized dark matter halos. We have varied the degree of magnetization as well as the halo mass ($1.8 \cdot 10^5$ -- $9.6 \cdot 10^7$ M$_{\sun}$) in order to infer systematical effects on the evolution of the magnetic field. The simulations were performed using a tabulated cooling function in order to model the thermodynamical behaviour of the gas. Despite the simplified physical treatment, our simulations agree well with results from other authors and thus serve as a useful guide for more realistic simulations.

Our simulations suggest that supernovae with a total explosion energy of 10$^{51}$/10$^{53}$ erg are not able to disrupt halos more massive than 10$^6$ M$_{\sun}$/10$^7$ M$_{\sun}$ in agreement with other studies~\citep[e.g.][]{Bromm03,Kitayama05,Whalen08,Ritter12}. In case the supernova manages to disrupt the halo, a thin shock layer develops separating the post-shock from the pre-shock material. The initially small-scale, strongly disordered magnetic field is transported with the expanding gas resulting in a significant increase in the coherence length of the gas. The magnetic field spectra show a shift of the peak down to $k$-values of about 4 at the end of the simulations corresponding to a length scale of up to 250 pc. The typical correlation length of the magnetic field determined from autocorrelation functions at the end of the simulations is of the order of 150 -- 200 pc in good agreement with the magnetic field spectra. We note that the maximum reachable coherence length in this work was limited by the size of the simulation domain. For gas with a reduced cooling efficiency a thicker post-shock region builds up and the spectra peak at larger $k$-values.

In general we find that for halos being disrupted by a supernova the large-scale component of the magnetic field has a strength ranging from about $10^{-12}$ G to $10^{-8}$ G strongly depending on the initial magnetic field strength. Moreover, the coherence length $l_\rmn{B}$ of the magnetic field scales approximately linear with the extension of the supernova bubble being about 20\% -- 50\% of the supernova shell radius. Extrapolating this relation for later times we find that for a typical PISN reaching an extension of up to 3 kpc \citep{Greif07} well-ordered magnetic fields with coherence lengths of up to 1.5 kpc can be expected. We note that due to the simplified treatment of the gas cooling and the initial modelling of the HII region and supernova blast wave, the above values should rather be considered as upper limits.

We discussed a number of implications of our findings for several subsequent processes. We suggest that magnetic fields with this strength and coherence length can have significant implications for a subsequent recollapse of the halo in particular by suppressing fragmentation \citep{Machida13}. Furthermore, such fields could potentially affect regions of a high halo density where several star forming halos with supernova explosions are expected, e.g. in the regions where the first galaxies are believed to form. Our simulations suggest that these regions might have a significant large-scale magnetic field component from the very beginning. This could affect the formation of first galaxies and the onset of the mean-field dynamo responsible for the large-scale magnetic field observed in galaxies nowadays. Finally, the occurrence of supernovae might have also contributed to the magnetization of the intergalactic medium already around a redshift of z = 20.

\section*{Acknowledgements}

The authors like to thank the anonymous referee for the very thorough report which helped to significantly improve the paper. The authors also like to thank S. Bovino and M. Latif for stimulating discussions. We acknowledges funding by the Deutsche Forschungsgemeinschaft via grants BA 3706/3-1 and SCHL 1964/1-1 within the SPP \textit{The interstellar medium}. DRGS also thanks for funding from the German Science Foundation via the SFB 963/1 "Astrophysical Flow Instabilities and Turbulence" (project A12). The simulations presented here were performed on the Supercomputer JUQUEEN at the Supercomputing Centre in J\"ulich. The FLASH code was partly developed by the DOE-supported Alliances Center for Astrophysical Thermonuclear Flashes (ASC) at the University of Chicago.


\begin{thebibliography}{87}
\expandafter\ifx\csname natexlab\endcsname\relax\def\natexlab#1{#1}\fi

\bibitem[{{Abel} {et~al.}(2000){Abel}, {Bryan}, \& {Norman}}]{Abel00}
{Abel}, T., {Bryan}, G.~L., \& {Norman}, M.~L. 2000, \apj, 540, 39

\bibitem[{{Abel} {et~al.}(2002){Abel}, {Bryan}, \& {Norman}}]{Abel02}
{Abel}, T., {Bryan}, G.~L., \& {Norman}, M.~L. 2002, Science, 295, 93

\bibitem[{{Abel} {et~al.}(2007){Abel}, {Wise}, \& {Bryan}}]{Abel07}
{Abel}, T., {Wise}, J.~H., \& {Bryan}, G.~L. 2007, \apjl, 659, L87

\bibitem[{{Arshakian} {et~al.}(2009){Arshakian}, {Beck}, {Krause}, \&
  {Sokoloff}}]{Arshakian09}
{Arshakian}, T.~G., {Beck}, R., {Krause}, M., \& {Sokoloff}, D. 2009, \aap,
  494, 21

\bibitem[{{Balsara} {et~al.}(2001){Balsara}, {Benjamin}, \& {Cox}}]{Balsara01}
{Balsara}, D., {Benjamin}, R.~A., \& {Cox}, D.~P. 2001, \apj, 563, 800

\bibitem[{{Banerjee} \& {Jedamzik}(2004)}]{Banerjee04}
{Banerjee}, R. \& {Jedamzik}, K. 2004, \prd, 70, 123003

\bibitem[{{Beck}(2012)}]{Beck12}
{Beck}, R. 2012, \ssr, 166, 215

\bibitem[{{Beck} {et~al.}(1999){Beck}, {Ehle}, {Shoutenkov}, {Shukurov}, \&
  {Sokoloff}}]{Beck99}
{Beck}, R., {Ehle}, M., {Shoutenkov}, V., {Shukurov}, A., \& {Sokoloff}, D.
  1999, \nat, 397, 324

\bibitem[{{Beck} {et~al.}(1994){Beck}, {Poezd}, {Shukurov}, \&
  {Sokoloff}}]{Beck94}
{Beck}, R., {Poezd}, A.~D., {Shukurov}, A., \& {Sokoloff}, D.~D. 1994, \aap,
  289, 94

\bibitem[{{Biermann}(1950)}]{Biermann50}
{Biermann}, L. 1950, Zeitschrift Naturforschung Teil A, 5, 65

\bibitem[{{Bromm} {et~al.}(1999){Bromm}, {Coppi}, \& {Larson}}]{Bromm99}
{Bromm}, V., {Coppi}, P.~S., \& {Larson}, R.~B. 1999, \apjl, 527, L5

\bibitem[{{Bromm} {et~al.}(2002){Bromm}, {Coppi}, \& {Larson}}]{Bromm02}
{Bromm}, V., {Coppi}, P.~S., \& {Larson}, R.~B. 2002, \apj, 564, 23

\bibitem[{{Bromm} {et~al.}(2003){Bromm}, {Yoshida}, \& {Hernquist}}]{Bromm03}
{Bromm}, V., {Yoshida}, N., \& {Hernquist}, L. 2003, \apjl, 596, L135

\bibitem[{{Chevalier} {et~al.}(1992){Chevalier}, {Blondin}, \&
  {Emmering}}]{Chevalier92}
{Chevalier}, R.~A., {Blondin}, J.~M., \& {Emmering}, R.~T. 1992, \apj, 392, 118

\bibitem[{{Clark} {et~al.}(2011){Clark}, {Glover}, {Klessen}, \&
  {Bromm}}]{Clark11}
{Clark}, P.~C., {Glover}, S.~C.~O., {Klessen}, R.~S., \& {Bromm}, V. 2011,
  \apj, 727, 110

\bibitem[{{de Souza} \& {Opher}(2010)}]{Souza10}
{de Souza}, R.~S. \& {Opher}, R. 2010, \prd, 81, 067301

\bibitem[{{de Souza} {et~al.}(2011){de Souza}, {Rodrigues}, {Ishida}, \&
  {Opher}}]{Souza11}
{de Souza}, R.~S., {Rodrigues}, L.~F.~S., {Ishida}, E.~E.~O., \& {Opher}, R.
  2011, \mnras, 415, 2969

\bibitem[{{Dolag} {et~al.}(2011){Dolag}, {Kachelriess}, {Ostapchenko}, \&
  {Tom{\`a}s}}]{Dolag11}
{Dolag}, K., {Kachelriess}, M., {Ostapchenko}, S., \& {Tom{\`a}s}, R. 2011,
  \apjl, 727, L4

\bibitem[{{Dubey} {et~al.}(2008){Dubey}, {Fisher}, {Graziani}, {Jordan},
  {Lamb}, {Reid}, {Rich}, {Sheeler}, {Townsley}, \& {Weide}}]{Dubey08}
{Dubey}, A., {Fisher}, R., {Graziani}, C., {et~al.} 2008, in Astronomical
  Society of the Pacific Conference Series, Vol. 385, Numerical Modeling of
  Space Plasma Flows, ed. N.~V. {Pogorelov}, E.~{Audit}, \& G.~P. {Zank}, 145

\bibitem[{{Eke} {et~al.}(1996){Eke}, {Cole}, \& {Frenk}}]{Eke96}
{Eke}, V.~R., {Cole}, S., \& {Frenk}, C.~S. 1996, \mnras, 282, 263

\bibitem[{{Fermi}(1949)}]{Fermi49}
{Fermi}, E. 1949, Physical Review, 75, 1169

\bibitem[{Fryxell {et~al.}(2000)Fryxell, Olson, Ricker, Timmes, Zingale, Lamb,
  MacNeice, Rosner, Truran, \& Tufo}]{Fryxell00}
Fryxell, B., Olson, K., Ricker, P., {et~al.} 2000, Astrophysical Journal,
  Supplement, 131, 273

\bibitem[{{Gao} {et~al.}(2010){Gao}, {Theuns}, {Frenk}, {Jenkins}, {Helly},
  {Navarro}, {Springel}, \& {White}}]{Gao10}
{Gao}, L., {Theuns}, T., {Frenk}, C.~S., {et~al.} 2010, \mnras, 403, 1283

\bibitem[{{Gao} {et~al.}(2007){Gao}, {Yoshida}, {Abel}, {Frenk}, {Jenkins}, \&
  {Springel}}]{Gao07}
{Gao}, L., {Yoshida}, N., {Abel}, T., {et~al.} 2007, \mnras, 378, 449

\bibitem[{{Gnedin} \& {Hollon}(2012)}]{Gnedin12}
{Gnedin}, N.~Y. \& {Hollon}, N. 2012, \apjs, 202, 13

\bibitem[{{Grassi} {et~al.}(2013){Grassi}, {Bovino}, {Schleicher}, {Prieto},
  {Seifried}, {Simoncini}, \& {Gianturco}}]{Grassi13}
{Grassi}, T., {Bovino}, S., {Schleicher}, D.~R.~G., {et~al.} 2013,
  arXiv:1311.1070

\bibitem[{{Greif} {et~al.}(2012){Greif}, {Bromm}, {Clark}, {Glover}, {Smith},
  {Klessen}, {Yoshida}, \& {Springel}}]{Greif12}
{Greif}, T.~H., {Bromm}, V., {Clark}, P.~C., {et~al.} 2012, \mnras, 424, 399

\bibitem[{{Greif} {et~al.}(2007){Greif}, {Johnson}, {Bromm}, \&
  {Klessen}}]{Greif07}
{Greif}, T.~H., {Johnson}, J.~L., {Bromm}, V., \& {Klessen}, R.~S. 2007, \apj,
  670, 1

\bibitem[{{Greif} {et~al.}(2008){Greif}, {Johnson}, {Klessen}, \&
  {Bromm}}]{Greif08}
{Greif}, T.~H., {Johnson}, J.~L., {Klessen}, R.~S., \& {Bromm}, V. 2008,
  \mnras, 387, 1021

\bibitem[{{Greif} {et~al.}(2011){Greif}, {Springel}, {White}, {Glover},
  {Clark}, {Smith}, {Klessen}, \& {Bromm}}]{Greif11}
{Greif}, T.~H., {Springel}, V., {White}, S.~D.~M., {et~al.} 2011, \apj, 737, 75

\bibitem[{{Heger} \& {Woosley}(2002)}]{Heger02}
{Heger}, A. \& {Woosley}, S.~E. 2002, \apj, 567, 532

\bibitem[{{Hirano} {et~al.}(2014){Hirano}, {Hosokawa}, {Yoshida}, {Umeda},
  {Omukai}, {Chiaki}, \& {Yorke}}]{Hirano14}
{Hirano}, S., {Hosokawa}, T., {Yoshida}, N., {et~al.} 2014, \apj, 781, 60

\bibitem[{{Hosokawa} {et~al.}(2011){Hosokawa}, {Omukai}, {Yoshida}, \&
  {Yorke}}]{Hosokawa11}
{Hosokawa}, T., {Omukai}, K., {Yoshida}, N., \& {Yorke}, H.~W. 2011, Science,
  334, 1250

\bibitem[{{Johnson} {et~al.}(2013){Johnson}, {Whalen}, {Even}, {Fryer},
  {Heger}, {Smidt}, \& {Chen}}]{Johnson13}
{Johnson}, J.~L., {Whalen}, D.~J., {Even}, W., {et~al.} 2013, \apj, 775, 107

\bibitem[{{Kazantsev}(1968)}]{Kazantsev68}
{Kazantsev}, A.~P. 1968, Soviet Journal of Experimental and Theoretical
  Physics, 26, 1031

\bibitem[{{Kim} {et~al.}(1989){Kim}, {Kronberg}, {Giovannini}, \&
  {Venturi}}]{Kim89}
{Kim}, K.-T., {Kronberg}, P.~P., {Giovannini}, G., \& {Venturi}, T. 1989, \nat,
  341, 720

\bibitem[{{Kitayama} \& {Yoshida}(2005)}]{Kitayama05}
{Kitayama}, T. \& {Yoshida}, N. 2005, \apj, 630, 675

\bibitem[{{Kitayama} {et~al.}(2004){Kitayama}, {Yoshida}, {Susa}, \&
  {Umemura}}]{Kitayama04}
{Kitayama}, T., {Yoshida}, N., {Susa}, H., \& {Umemura}, M. 2004, \apj, 613,
  631

\bibitem[{{Kronberg}(1994)}]{Kronberg94}
{Kronberg}, P.~P. 1994, Reports on Progress in Physics, 57, 325

\bibitem[{{Kronberg} {et~al.}(1999){Kronberg}, {Lesch}, \& {Hopp}}]{Kronberg99}
{Kronberg}, P.~P., {Lesch}, H., \& {Hopp}, U. 1999, \apj, 511, 56

\bibitem[{{Kulsrud} \& {Zweibel}(2008)}]{Kulsrud08}
{Kulsrud}, R.~M. \& {Zweibel}, E.~G. 2008, Reports on Progress in Physics, 71,
  046901

\bibitem[{{Latif} {et~al.}(2013{\natexlab{a}}){Latif}, {Schleicher}, \&
  {Schmidt}}]{Latif13d}
{Latif}, M.~A., {Schleicher}, D.~R.~G., \& {Schmidt}, W. 2013{\natexlab{a}},
  arXiv:1310.3680

\bibitem[{{Latif} {et~al.}(2013{\natexlab{b}}){Latif}, {Schleicher}, {Schmidt},
  \& {Niemeyer}}]{Latif13a}
{Latif}, M.~A., {Schleicher}, D.~R.~G., {Schmidt}, W., \& {Niemeyer}, J.
  2013{\natexlab{b}}, \mnras, 430, 588

\bibitem[{{Latif} {et~al.}(2013{\natexlab{c}}){Latif}, {Schleicher}, {Schmidt},
  \& {Niemeyer}}]{Latif13c}
{Latif}, M.~A., {Schleicher}, D.~R.~G., {Schmidt}, W., \& {Niemeyer}, J.
  2013{\natexlab{c}}, \apjl, 772, L3

\bibitem[{{Latif} {et~al.}(2013{\natexlab{d}}){Latif}, {Schleicher}, {Schmidt},
  \& {Niemeyer}}]{Latif13b}
{Latif}, M.~A., {Schleicher}, D.~R.~G., {Schmidt}, W., \& {Niemeyer}, J.
  2013{\natexlab{d}}, \mnras, 432, 668

\bibitem[{Lee \& Deane(2009)}]{Lee09}
Lee, D. \& Deane, A.~E. 2009, J. Comput. Phys., 228, 952

\bibitem[{{Machida} \& {Doi}(2013)}]{Machida13}
{Machida}, M.~N. \& {Doi}, K. 2013, \mnras, 435, 3283

\bibitem[{{Mouschovias} \& {Spitzer}(1976)}]{Mouschovias76}
{Mouschovias}, T.~C. \& {Spitzer}, Jr., L. 1976, \apj, 210, 326

\bibitem[{{Navarro} {et~al.}(1997){Navarro}, {Frenk}, \& {White}}]{Navarro97}
{Navarro}, J.~F., {Frenk}, C.~S., \& {White}, S.~D.~M. 1997, \apj, 490, 493

\bibitem[{{Neronov} {et~al.}(2013){Neronov}, {Semikoz}, \&
  {Banafsheh}}]{Neronov13}
{Neronov}, A., {Semikoz}, D., \& {Banafsheh}, M. 2013, arXiv:1305.1450

\bibitem[{{Neronov} \& {Vovk}(2010)}]{Neronov10}
{Neronov}, A. \& {Vovk}, I. 2010, Science, 328, 73

\bibitem[{{O'Shea} \& {Norman}(2007)}]{OShea07}
{O'Shea}, B.~W. \& {Norman}, M.~L. 2007, \apj, 654, 66

\bibitem[{{O'Shea} \& {Norman}(2008)}]{OShea08}
{O'Shea}, B.~W. \& {Norman}, M.~L. 2008, \apj, 673, 14

\bibitem[{{Ostriker} \& {McKee}(1988)}]{Ostriker88}
{Ostriker}, J.~P. \& {McKee}, C.~F. 1988, Reviews of Modern Physics, 60, 1

\bibitem[{{Parker}(1955)}]{Parker55}
{Parker}, E.~N. 1955, \apj, 122, 293

\bibitem[{{Peters} {et~al.}(2012){Peters}, {Schleicher}, {Klessen}, {Banerjee},
  {Federrath}, {Smith}, \& {Sur}}]{Peters12}
{Peters}, T., {Schleicher}, D.~R.~G., {Klessen}, R.~S., {et~al.} 2012, \apjl,
  760, L28

\bibitem[{{Press} \& {Schechter}(1974)}]{Press74}
{Press}, W.~H. \& {Schechter}, P. 1974, \apj, 187, 425

\bibitem[{{Ritter} {et~al.}(2012){Ritter}, {Safranek-Shrader}, {Gnat},
  {Milosavljevi{\'c}}, \& {Bromm}}]{Ritter12}
{Ritter}, J.~S., {Safranek-Shrader}, C., {Gnat}, O., {Milosavljevi{\'c}}, M.,
  \& {Bromm}, V. 2012, \apj, 761, 56

\bibitem[{{Ruzmaikin} {et~al.}(1988){Ruzmaikin}, {Sokolov}, \&
  {Shukurov}}]{Ruzmaikin88}
{Ruzmaikin}, A.~A., {Sokolov}, D.~D., \& {Shukurov}, A.~M., eds. 1988,
  Astrophysics and Space Science Library, Vol. 133, {Magnetic fields of
  galaxies}

\bibitem[{{Schleicher} {et~al.}(2010){Schleicher}, {Banerjee}, {Sur},
  {Arshakian}, {Klessen}, {Beck}, \& {Spaans}}]{Schleicher10}
{Schleicher}, D.~R.~G., {Banerjee}, R., {Sur}, S., {et~al.} 2010, \aap, 522,
  A115

\bibitem[{{Schleicher} {et~al.}(2013){Schleicher}, {Schober}, {Federrath},
  {Bovino}, \& {Schmidt}}]{Schleicher13}
{Schleicher}, D.~R.~G., {Schober}, J., {Federrath}, C., {Bovino}, S., \&
  {Schmidt}, W. 2013, New Journal of Physics, 15, 023017

\bibitem[{{Schlickeiser}(2012)}]{Schlickeiser12}
{Schlickeiser}, R. 2012, Physical Review Letters, 109, 261101

\bibitem[{{Schober} {et~al.}(2012{\natexlab{a}}){Schober}, {Schleicher},
  {Federrath}, {Glover}, {Klessen}, \& {Banerjee}}]{Schober12b}
{Schober}, J., {Schleicher}, D., {Federrath}, C., {et~al.} 2012{\natexlab{a}},
  \apj, 754, 99

\bibitem[{{Schober} {et~al.}(2012{\natexlab{b}}){Schober}, {Schleicher},
  {Federrath}, {Klessen}, \& {Banerjee}}]{Schober12a}
{Schober}, J., {Schleicher}, D., {Federrath}, C., {Klessen}, R., \& {Banerjee},
  R. 2012{\natexlab{b}}, \pre, 85, 026303

\bibitem[{{Schober} {et~al.}(2013){Schober}, {Schleicher}, \&
  {Klessen}}]{Schober13}
{Schober}, J., {Schleicher}, D.~R.~G., \& {Klessen}, R.~S. 2013, \aap, 560, A87

\bibitem[{{Sigl} {et~al.}(1997){Sigl}, {Olinto}, \& {Jedamzik}}]{Sigl97}
{Sigl}, G., {Olinto}, A.~V., \& {Jedamzik}, K. 1997, \prd, 55, 4582

\bibitem[{{Stacy} \& {Bromm}(2013)}]{Stacy13}
{Stacy}, A. \& {Bromm}, V. 2013, \mnras, 433, 1094

\bibitem[{{Subramanian}(1997)}]{Subramanian97}
{Subramanian}, K. 1997, arXiv:astro-ph/9708216

\bibitem[{{Sur} {et~al.}(2012){Sur}, {Federrath}, {Schleicher}, {Banerjee}, \&
  {Klessen}}]{Sur12}
{Sur}, S., {Federrath}, C., {Schleicher}, D.~R.~G., {Banerjee}, R., \&
  {Klessen}, R.~S. 2012, \mnras, 423, 3148

\bibitem[{{Sur} {et~al.}(2010){Sur}, {Schleicher}, {Banerjee}, {Federrath}, \&
  {Klessen}}]{Sur10}
{Sur}, S., {Schleicher}, D.~R.~G., {Banerjee}, R., {Federrath}, C., \&
  {Klessen}, R.~S. 2010, \apjl, 721, L134

\bibitem[{{Susa}(2013)}]{Susa13}
{Susa}, H. 2013, \apj, 773, 185

\bibitem[{{Sutherland} \& {Dopita}(1993)}]{Sutherland93}
{Sutherland}, R.~S. \& {Dopita}, M.~A. 1993, \apjs, 88, 253

\bibitem[{{Takahashi} {et~al.}(2012){Takahashi}, {Mori}, {Ichiki}, \&
  {Inoue}}]{Takahashi12}
{Takahashi}, K., {Mori}, M., {Ichiki}, K., \& {Inoue}, S. 2012, \apjl, 744, L7

\bibitem[{{Turk} {et~al.}(2009){Turk}, {Abel}, \& {O'Shea}}]{Turk09}
{Turk}, M.~J., {Abel}, T., \& {O'Shea}, B. 2009, Science, 325, 601

\bibitem[{{Turk} {et~al.}(2012){Turk}, {Oishi}, {Abel}, \& {Bryan}}]{Turk12}
{Turk}, M.~J., {Oishi}, J.~S., {Abel}, T., \& {Bryan}, G.~L. 2012, \apj, 745,
  154

\bibitem[{{Turner} \& {Widrow}(1988)}]{Turner88}
{Turner}, M.~S. \& {Widrow}, L.~M. 1988, \prd, 37, 2743

\bibitem[{{Vishniac}(1983)}]{Vishniac83}
{Vishniac}, E.~T. 1983, \apj, 274, 152

\bibitem[{{Wagstaff} {et~al.}(2013){Wagstaff}, {Banerjee}, {Schleicher}, \&
  {Sigl}}]{Wagstaff13}
{Wagstaff}, J.~M., {Banerjee}, R., {Schleicher}, D., \& {Sigl}, G. 2013,
  arXiv:1304.4723

\bibitem[{{Whalen} {et~al.}(2004){Whalen}, {Abel}, \& {Norman}}]{Whalen04}
{Whalen}, D., {Abel}, T., \& {Norman}, M.~L. 2004, \apj, 610, 14

\bibitem[{{Whalen} \& {Norman}(2006)}]{Whalen06}
{Whalen}, D. \& {Norman}, M.~L. 2006, \apjs, 162, 281

\bibitem[{{Whalen} {et~al.}(2008){Whalen}, {van Veelen}, {O'Shea}, \&
  {Norman}}]{Whalen08}
{Whalen}, D., {van Veelen}, B., {O'Shea}, B.~W., \& {Norman}, M.~L. 2008, \apj,
  682, 49

\bibitem[{{Whalen} {et~al.}(2013{\natexlab{a}}){Whalen}, {Johnson}, {Smidt},
  {Heger}, {Even}, \& {Fryer}}]{Whalen13b}
{Whalen}, D.~J., {Johnson}, J.~L., {Smidt}, J., {et~al.} 2013{\natexlab{a}},
  \apj, 777, 99

\bibitem[{{Whalen} {et~al.}(2013{\natexlab{b}}){Whalen}, {Johnson}, {Smidt},
  {Meiksin}, {Heger}, {Even}, \& {Fryer}}]{Whalen13a}
{Whalen}, D.~J., {Johnson}, J.~L., {Smidt}, J., {et~al.} 2013{\natexlab{b}},
  \apj, 774, 64

\bibitem[{{Wise} \& {Abel}(2008)}]{Wise08}
{Wise}, J.~H. \& {Abel}, T. 2008, \apj, 685, 40

\bibitem[{{Yoshida} {et~al.}(2003){Yoshida}, {Abel}, {Hernquist}, \&
  {Sugiyama}}]{Yoshida03}
{Yoshida}, N., {Abel}, T., {Hernquist}, L., \& {Sugiyama}, N. 2003, \apj, 592,
  645

\bibitem[{{Yoshida} {et~al.}(2007){Yoshida}, {Oh}, {Kitayama}, \&
  {Hernquist}}]{Yoshida07}
{Yoshida}, N., {Oh}, S.~P., {Kitayama}, T., \& {Hernquist}, L. 2007, \apj, 663,
  687

\bibitem[{{Yoshida} {et~al.}(2006){Yoshida}, {Omukai}, {Hernquist}, \&
  {Abel}}]{Yoshida06}
{Yoshida}, N., {Omukai}, K., {Hernquist}, L., \& {Abel}, T. 2006, \apj, 652, 6

\end{thebibliography}

\label{lastpage}

\end{document}